\documentclass[manuscript,screen,authorversion,nonacm]{acmart}

\AtBeginDocument{%
  \providecommand\BibTeX{{%
    \normalfont B\kern-0.5em{\scshape i\kern-0.25em b}\kern-0.8em\TeX}}}

\setcopyright{rightsretained}
\copyrightyear{2023}
\acmYear{2023}

\usepackage{tikz}

\tikzset{every picture/.style={line width=0.75pt}} 

\usepackage{multirow}
\usepackage{subcaption}
\usepackage{pgfplots}
\pgfplotsset{compat=1.14} 
\usepackage[font=small,skip=5pt]{caption}
\usepackage[font=small,skip=-12pt]{subcaption}

\def \plotheight {4.22cm}
\def \plotwidth {8cm}

\newcommand{\saverio}[1]{{\color{black} #1}}

\begin{document}

\title{A Classification of Feedback Loops and Their Relation to Biases in Automated Decision-Making Systems}

\author{Nicol\`{o} Pagan}
\authornote{Both authors contributed equally to this research.}
\email{nicolo.pagan@uzh.ch}
\orcid{0000-0003-0071-7856}
\affiliation{%
  \institution{University of Zurich}
  \city{Zurich}
  \country{Switzerland}
}

\author{Joachim Baumann}
\authornotemark[1]
\email{baumann@ifi.uzh.ch}
\orcid{0000-0003-2019-4829}
\affiliation{%
  \institution{University of Zurich}
  \city{Zurich}
  \country{Switzerland}
}
\affiliation{%
  \institution{Zurich University of Applied Sciences}
  \city{Zurich}
  \country{Switzerland}
}

\author{Ezzat Elokda}
\email{elokdae@ethz.ch}
\orcid{0000-0002-2415-6494}
\affiliation{%
  \institution{ETH Zurich}
  \city{Zurich}
  \country{Switzerland}
}

\author{Giulia De Pasquale}
\email{degiulia@ethz.ch}
\orcid{0000-0002-8587-967X}
\affiliation{%
  \institution{ETH Zurich}
  \city{Zurich}
  \country{Switzerland}
}

\author{Saverio Bolognani}
\email{bsaverio@ethz.ch}
\orcid{0000-0002-7935-1385}
\affiliation{%
  \institution{ETH Zurich}
  \city{Zurich}
  \country{Switzerland}
}

\author{Anik\'{o} Hann\'{a}k}
\email{hannak@ifi.uzh.ch}
\orcid{0000-0002-0612-6320}
\affiliation{%
  \institution{University of Zurich}
  \city{Zurich}
  \country{Switzerland}
}

\renewcommand{\shortauthors}{Pagan and Baumann, et al.}

\begin{abstract}
Prediction-based decision-making systems are becoming increasingly prevalent in various domains. Previous studies have demonstrated that such systems are vulnerable to runaway feedback loops, e.g., when police are repeatedly sent back to the same neighborhoods regardless of the actual rate of criminal activity, which exacerbate existing biases. 
In practice, the automated decisions have dynamic feedback effects on the system itself that can perpetuate over time, making it difficult for short-sighted design choices to control the system's evolution.
While researchers started proposing longer-term solutions to prevent adverse outcomes (such as bias towards certain groups), these interventions largely depend on ad hoc modeling assumptions and a rigorous theoretical understanding of the feedback dynamics in ML-based decision-making systems is currently missing.
In this paper, we use the language of dynamical systems theory, a branch of applied mathematics that deals with the analysis of the interconnection of systems with dynamic behaviors, to rigorously classify the different types of feedback loops in the ML-based decision-making pipeline.
By reviewing existing scholarly work, we show that this classification covers many examples discussed in the algorithmic fairness community, thereby providing a unifying and principled framework to study feedback loops. By qualitative analysis, and through a simulation example of recommender systems, we show which specific types of ML biases are affected by each type of feedback loop. We find that the existence of feedback loops in the ML-based decision-making pipeline can perpetuate, reinforce, or even reduce ML biases.
\end{abstract}


\maketitle

\section{Introduction}\label{sec:introduction}

Many of today's automated processes rely on machine learning (ML) algorithms to inform decisions that have a profound impact on people's lives. 
For instance, they are employed to evaluate whether an individual should be admitted to a certain college~\cite{Kleinberg2018}, be granted a loan~\cite{Fuster2022PredictablyMarkets}, or treated as high risk of recidivism~\cite{berk2021criminal}.
The advantage of these ML-based decision-making systems is their scalability, i.e., the capability to handle a vast number of decisions in an efficient manner.
However, researchers have found evidence that these algorithms often exacerbate existing biases that underlie human decisions~\cite{Grove2000,Dawes1989,Kleinberg2017HumanDecisionsMachinePredictions} and even introduce new ones~\cite{simonite2015probing,angwin2016machine,crawford2016artificial,pmlr-v81-buolamwini18a}.

To solve this problem, a recent line of research in algorithmic fairness started investigating solutions that can mitigate these biases by enforcing some metrics of individual or group fairness~\mbox{\cite{caton2020fairness,mehrabi2021survey,10.1145/3097983.3098095,hardt2016equality,baumann2022sufficiency}}.
Although these attempts prove to be successful in the short term, 
they often do not perform equally well in the long term, i.e., after multiple rounds of the decision-making process~\cite{Liu2018Delayed_Long_version, sunbackfire}.%
\footnote{More notably, enforcing fairness constraints often leads to counter-intuitive and undesired results, increasing the gap between advantaged and disadvantaged groups, thus again exacerbating existing initial biases~\cite{DAmour2020FairnessStudies}.}
The underlying reason seems to lie in the fact that the mitigating solutions are designed for stationary systems~\cite{Chouldechova2018,mitchell2021algorithmic}, while the system itself dynamically evolves over time.
More specifically, the system changes over time because the output (the decision) feeds back as input to the system itself, thus creating what researchers refer to as a ``feedback loop''.
The result is that biases are perpetuated (or even reinforced) due to the existence of the feedback loop, despite enforcing the mitigation techniques. Even though researchers recently started studying the long-term effects of sequential decision-making algorithms (e.g., \cite{DAmour2020FairnessStudies, Liu2018,Hu2018AMarket}, see~\cite{Zhang2021FairnessSurvey} for a recent survey), 
the proposed simulation-based solutions are drawn on ad hoc models which prevent a comparison of their underlying assumptions and a deep interpretation of the driving factors, i.e., what causes the feedback loops and which components of the system are involved.
As a result, to date, we lack a comprehensive classification and theoretical understanding of these feedback loops, and how they relate to the amplification of different types of bias.
This is a necessary first step to change the research perspective from developing short-sighted solutions aimed at identifying and fixing existing biases to a more long-term-oriented view that strives to anticipate and prevent biases.

Compared to previous work in the field, in this paper, we do not attempt to provide a simulation-based solution to the existence of feedback loops, rather we fill this theoretical gap by providing a formal definition and a rigorous taxonomy of feedback loops in the ML-based decision-making pipeline, and by linking them to the biases they affect.
To do so, we first clarify the difference between open-loop and closed-loop (or feedback-loop) systems by borrowing the language and the tools from 
dynamical systems theory, the discipline that focuses on the analysis of systems with dynamical behavior (and their interconnection).
Then, we apply this system-theoretic framework to the decision-making pipeline, which is composed of different sub-systems: the individuals' sampling process, the individuals' characteristics representing the decision-relevant construct, the observed features and outcomes, the ML model, and the final decision.
The final decision can feed back into any of these sub-systems, thus forming different types of feedback loops.
This, in turn, means that the ultimate effect on the whole pipeline and the amplification of biases depend on what types of feedback loops are simultaneously present in the system.

The first contribution of this paper 
(see Sec.~\ref{sec:preliminaries}) 
is to cast the ML-based decision-making pipeline into a
system-theoretic framework that emphasizes the different components.
Our second contribution 
(see Sec.~\ref{sec:A-Formal-Framework-for-Feedback-Loops})
consists in providing a classification of the different types of feedback loops, which we call \textit{sampling}, \textit{individual}, \textit{feature}, \textit{outcome}, and \textit{ML model feedback loop} depending on which component is affected. 
Additionally, we introduce the notion of ``adversarial feedback loops,'' which represent special cases of feedback loops in which the final decision feeds back into the system as a consequence of some strategic action of the affected individual(s).
As a third contribution
(see Sec.~\ref{sec:implications-algo-fairness}), 
we provide an overview of the different types of bias that can be affected by each of the five feedback loops we introduce.
As a fourth and final contribution
(see Sec.~\ref{sec:example}), we demonstrate the potential of our classification framework in the context of news recommender systems.
Specifically, we demonstrate that different types of feedback loops can affect distinct parts of the decision-making pipeline, resulting in system dynamics that produce various forms of bias.

\section{The ML-based decision-making pipeline in the frame of Dynamical Systems Theory}
\label{sec:preliminaries}

Accounting for the fact that ML-based decision-making systems are usually not static but evolve over time, we follow~\citet{dobbe2018broader} in using the language of dynamical systems theory to describe them.
A \emph{dynamical system} is a process that relates a set of \emph{input signals} to a set of \emph{output signals}.
A \emph{signal} is a variable or quantity of interest that may vary over time.
Thus, an algorithm is an example of a dynamical system that receives observable features as input signals and produces predictions or decisions as output signals.
Dynamical systems theory is concerned with the mathematical modeling of dynamical systems with the objective of understanding and/or manipulating fundamental properties, such as whether the system reaches a predictable operating point or exhibits oscillatory behaviors.

It is common to represent dynamical systems in block diagrams, where blocks denote systems and arrows denote signals, as a way to provide a high-level graphical representation of a real-world system.
Block diagrams are particularly useful to understand and study the \emph{interconnection} of different (sub-)systems, which are composed to form larger systems.
A \emph{series interconnection} occurs when the output of a system (or algorithm) is the input for another one.
A \emph{parallel interconnection} occurs when the same input enters two systems whose outputs are then combined.
In a \textit{feedback interconnection}, the output of a system is injected back as an input to one (or more) of  its components, creating a \emph{feedback loop}.
Series and parallel interconnections lead to \emph{open-loop systems}, whereas feedback interconnections lead to \emph{closed-loop systems}
-- see Fig.~\ref{fig:systems-interconnection} in Appendix~\ref{app:Open_and_closed_loop_dynamical_systems} for a visual representation.

\begin{figure}[t]
\includegraphics[width=.9\textwidth]{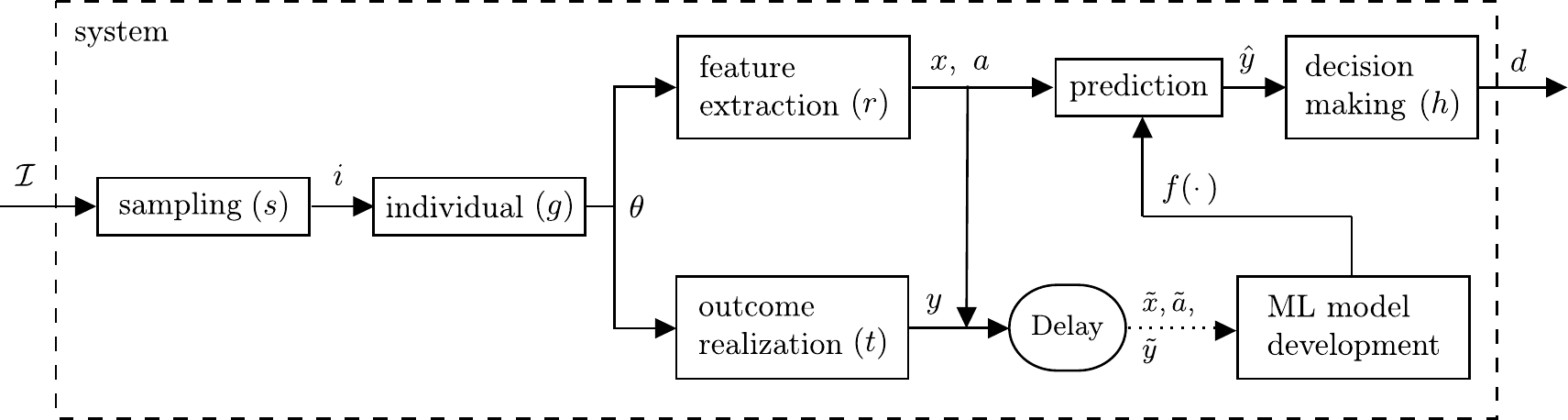}
\centering
\caption{The ML-based decision-making pipeline as an open-loop system.}
\label{fig:DM_in_control}
\end{figure}

The prototypical ML-based decision-making pipeline can also be represented as a block diagram.
We start by describing its open-loop components, shown in Fig.~\ref{fig:DM_in_control}, before characterizing possible feedback interconnections in Section~\ref{sec:A-Formal-Framework-for-Feedback-Loops}.
At the beginning of the pipeline, an individual $i$ is sampled from the world (i.e., the environment) $\mathcal{I}$, which represents a signal entering in the sampling function block $s:\mathcal{I} \rightarrow i$.
Let $i$ be the individual's identity -- i.e., its index in the population, which~\cite{Hertweck2021statistical-parity} call \textit{potential space} (PS) -- and let $g:i \rightarrow \theta$ be a function that returns the individual's attributes.
More precisely, $\theta$ denotes the construct that is relevant for the prediction -- what~\cite{Friedler2016} call \textit{construct space} (CS).
The features $x$, extracted through the function $r : \theta \rightarrow x,a$, and the outcome $y$ (also called label or target), realized through the function $t : \theta \rightarrow y$, are imperfect proxies that can be measured -- what~\cite{Friedler2016} call \textit{observed space} (OS).
For instance, $y$ can represent whether or not an individual repays a granted loan and $x$ is a set of features (for example, the credit score, as widely used in the US) that are used by the decision-maker to predict the repayment probability $\hat{y}$ in order to decide whether to grant the loan or not.
For each sampled individual, the final decision $d$ is informed by the prediction $\hat{y}$, which is produced based on the observed features $x$ to approximate $y$ using a learned function $f : x\rightarrow \hat{y}$.
Once the outcome is observed, i.e., after one time-unit of delay, the past time's feature label pair $\left(\tilde x,\tilde y\right)$ can end up as a sample in the dataset $(X,Y)$ that is used to (re)train and (re)evaluate an ML model (more details on the ML model development process are discussed in Appendix~\ref{app:detailed_ML_pipeline}).
In fully-automated decision-making systems, the decision rule $h$ is solely based on the prediction ($h : \hat{y} \rightarrow d$), usually taking the form of a simple threshold rule, e.g., $d=1$ if and only if $\hat{y}\geq \bar{y}$.
The symbol $a$ indicates the sensitive attribute of the individual (e.g., race or gender) and can possibly also be incorporated in the features $x$.
More precisely, the training, evaluation, prediction, or decision-making can use the information on the individual group memberships.%
\footnote{
Notice that $d$ does not always directly follow from $\hat{y}$.
Efforts to ensure group fairness usually take the group membership $a$ into account, e.g., to avoid disparate impact~\cite{pessach2022survey,caton2020fairness}.
Similarly, in non-automated decision-making systems, human decision-makers might consider any external, environmental information $z$, resulting in a more complex decision rule $h : f,x,a,\hat{y},z \rightarrow d$.
}

\section{Feedback Loops in the ML-based decision-making pipeline}
\label{sec:A-Formal-Framework-for-Feedback-Loops}

In contrast to ML, in the field of dynamical systems theory, feedback loops are not always seen as an undesirable feature of a system.
Lots of the emphasis of dynamical systems theory is on relating properties of the open-loop system, i.e., the system without a feedback loop, to those of the closed-loop system, i.e., the system with a feedback loop.
In this paper, we leverage the idea of closed-loop system properties to define feedback mechanisms in ML-based decision-making systems.
Interestingly, closed-loop systems may exhibit desirable properties compared to their open-loop counterparts.

In this section, we complete the specification of the ML pipeline as a dynamical system by considering the feedback interconnections that could be present.
We first define various types of feedback loops depending on the component of the ML pipeline affected by the outcome of the system (i.e., the final decision of the decision-maker).
Next, we introduce the concept of \textit{adversarial feedback loops}.
Then, we describe how different types of feedback loops can coexist.
Finally, we clarify some terminology with respect to positive and negative feedback loops.

\subsection{Feedback Loops}\label{ssec:Feedback-loops}
\begin{figure}[ht]
\includegraphics[width=.9\textwidth]{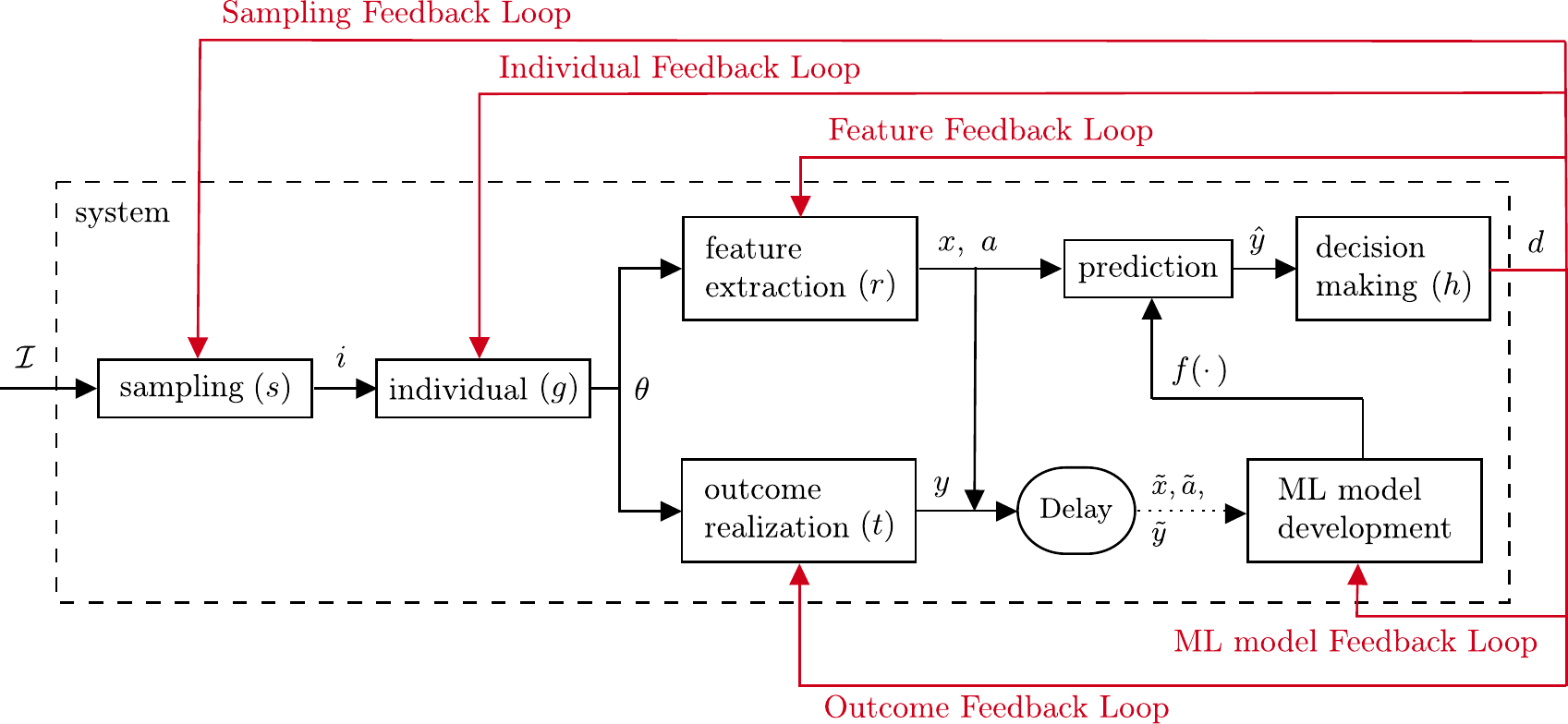}
\centering
\caption{The ML-based decision-making pipeline as a closed-loop system in which different feedback loops can emerge.}
\label{fig:feedback-loops}
\end{figure}

In many real-life settings, the decision taken at the end of the ML pipeline may feed back into some of its blocks.
Every block in the ML pipeline (except the prediction block, as this usually simply consists of applying $f$ to a new input example $x$) can be affected by the decision, each forming a different type of feedback loop, as depicted in Fig.~\ref{fig:feedback-loops}.
In what follows, we classify these feedback loops to provide a vocabulary and some examples.
To validate this terminology, we reviewed a total of 24 recent relevant papers that discuss issues of feedback loops in the context of ML-based decision-making systems -- many of which particularly focus on fairness aspects.
These papers are listed in Table~\ref{table:literature} (we describe the literature search process in more detail in Appendix~\ref{app:literature_search}).
We emphasize that the classification of the five feedback loops represented in Fig.~\ref{fig:feedback-loops} is complete with respect to the examples and use cases we identified in the current state of the literature on fair dynamic decision-making systems.
Despite covering existing literature, this feedback loop classification can easily be extended to capture more nuanced kinds of feedback.%
\footnote{
Notice that certain blocks of the ML pipeline depicted in Fig.~\ref{fig:feedback-loops} aggregate several processes that could be split up into several blocks, potentially resulting in a more nuanced classification of feedback loops.
For example, the block denominated ``feature extraction'' could be split up into measurement followed by feature engineering, creating two subcategories of the feature feedback loop.
However, existing works modeling feature feedback loops (e.g.,~\cite{Liu2018Delayed_Long_version,Milli2019}) consider the effect of the decision on the distribution of $x,a$, which forms the input for the prediction model, without differentiating between the effects of the decision on the measurement and the engineered features.
Similarly, we consider the entire ML model development process as one block with (one or more) new feature-label pairs as input and a learned prediction function as output (see Appendix~\ref{app:detailed_ML_pipeline} for more details).
}

\begin{table*}[]
\centering
\caption{Overview of feedback loops in the algorithmic fairness literature}
\label{table:literature}
\begin{tabular}{lll}
\toprule
\textbf{Feedback loop} &
  \textit{\textbf{non-adversarial}} &
  \textit{\textbf{adversarial}} \\
\midrule
Sampling Feedback Loop &
  \cite{
  pmlr-v80-hashimoto18a,Zhang2019GroupFairness,Zhang2020Long-TermLearning} & --
   \\ \hline
Individual Feedback Loop &
  \cite{9516926, perra2019modelling} &
  \cite{Hu2018AMarket, DAmour2020FairnessStudies, Heidari2019OnLearning,Kleinberg2020HowStrategically,Liu2020thedisparate, Zhang2020HowQualification} \\ \hline
Feature Feedback Loop &
  \cite{chaney2018algorithmic, Liu2018Delayed_Long_version,Zhang2020Long-TermLearning,DAmour2020FairnessStudies,sunbackfire,AliciaYiSun2022PhDThesis, NIPS2016_962e56a8} &
  \cite{Hu2018TheManipulation,Heidari2019OnLearning,Tsirtsis2019OptimalBehavior,Milli2019,Kleinberg2020HowStrategically,DAmour2020FairnessStudies,Liu2020thedisparate,Perdomo2020PerformativePrediction} \\ \hline
ML Model Feedback Loop &
  \cite{NIPS2016_962e56a8,Ensign2018RunawayPolicing,Ensign2018DecisionPrediction,Bechavod2019EqualFeedback,Elzayn2019FairProblems} & --
   \\ \hline
Outcome Feedback Loop &
  \cite{Perdomo2020PerformativePrediction} & --\\
\bottomrule
\end{tabular}
\end{table*}

\subsubsection{Sampling Feedback Loop}
\label{sssec:Sampling_Feedback_Loop}
The first type of feedback loop we introduce is the one that comprises the effects of the decision on the sampling of the individual from the population.
This influences the retention rate of different groups and modifies their representation.
Consider the following example of a college admission scenario discussed in~\cite{Mouzannar2019FromEquality}. 
First, let the total population be partitioned into two groups $A$ and $B$.
The population undergoes a selection process in which an institution, the decision-maker, designs a policy that maps each individual to a probability of being selected, possibly depending on the group identity $a$ and on observable attributes $x$ that bear information about qualification, e.g., GPA, SAT, or recommendation letters.
According to the authors of~\cite{Mouzannar2019FromEquality}, the selection process at time $t$ might change the qualification profiles of either group at time $t+1$ through a self-selection process acting in the form of filtering the pool of individuals available at the next iteration.
In other words, with the existence of a \emph{sampling} feedback loop, individuals belonging to a group that had received lower admission rates at the previous iteration might be discouraged from applying as candidates at the next iteration, thus affecting the application rates from the two groups (and ultimately the selection rates).
Note that, this feedback loop might lead to one of the two groups disappearing from the candidate pool.
To understand this, consider a similar example related to speech recognition products such as Amazon's Alexa and Google Home, which have been shown to have accent bias against non-native speakers~\cite{AmazonAlexa2022}, with native speakers experiencing much higher quality than non-native speakers. 
This difference can lead to a sampling feedback loop, where non-native speakers cease to use such products.
This may be hard to detect because the speech recognition model, from that point on, only receives input and training data from native speakers, potentially resulting in a model that is even more skewed towards the remaining users, i.e., the native speakers.
Without intervention, the model becomes even less accurate for non-native speakers, which reinforces the initial user experience~\cite{pmlr-v80-hashimoto18a}.
Additional examples of the sampling feedback loop can also be found in \cite{Zhang2019GroupFairness,Zhang2020Long-TermLearning}.

\subsubsection{Individual Feedback Loop}
\label{sssec:Individual_Feedback_Loop}
Another possible effect of the decision acts directly on the individual's characteristics $\theta$, i.e., through the function $g$.
An example of this type of feedback loop can be found in the users' reactions to personalized recommendations. 
As discussed in~\cite{9516926, perra2019modelling}, a user's opinion on, e.g., a certain political issue, is influenced by the news articles received. 
Therefore, the decision of the recommender system to promote a certain type of content has the effect of shifting the opinion of the individuals that receive such a recommendation.
Additional examples of the individual feedback loop are discussed in the context of adversarial feedback loops (see Sec.~\ref{ssec:adversarial_feedback_loops}).

\subsubsection{Feature Feedback Loop}
\label{sssec:Feature_Feedback_Loop}
The third type of feedback loop is relatively close to the previous one. However, in contrast to the individual feedback loop, the decision has an effect on the {\it observable} characteristics of the individual rather than on the actual ones, i.e., on $x$ rather than $\theta$.
One of the most common examples of this feature feedback loop can be found in credit lending scenarios in which a lender decides whether or not to approve a loan application based on the applicant's credit score, which is interpreted as a measurable and observable proxy for the individual's capability of paying back a granted loan~\cite{Liu2018Delayed_Long_version}.
For any positive decision, we observe a feature feedback loop: if the loan is repaid, the credit score increases; otherwise, if the applicant defaults, the credit score decreases.
Note that, in this example, the feedback loop takes place only if the decision is positive, and it also requires information on the actual outcome $y$. However, none of these conditions is strictly necessary for a feature feedback loop to occur.%
\footnote{
Suppose granting a loan is already enough to increase an individual's credit score.
In this case, the outcome of the lender's decision is fed back into $x$, creating a feature feedback loop irrespective of the realized outcome $y$.
}

Another example is constituted by content recommender systems where the time a user looks at some content is part of the observation captured in the feature $x$ \cite{NIPS2016_962e56a8, chaney2018algorithmic}.
However, the time explicitly depends on what the recommender system has previously suggested, thus closing a feature feedback loop.
This happens irrespective of whether this recommendation affects the individual's interests, i.e., even in the absence of an individual feedback loop.

Additional examples of the feature feedback loop can also be found in \cite{Zhang2020Long-TermLearning,DAmour2020FairnessStudies,sunbackfire,AliciaYiSun2022PhDThesis}.
Furthermore, similarly to the individual feedback loop, also for the feature feedback loop, there exists an adversarial counterpart (see Sec.~\ref{ssec:adversarial_feedback_loops}).

\subsubsection{ML Model Feedback Loop}
\label{sssec:ML_Model_Feedback_Loop}
In the ML model feedback loop, the final decision $d$ affects the ML model by modifying the training or the validation data sets $\left(X,Y\right)$ that will be used for future predictions.
Typical examples in this category are known as ML-based decision-making with \textit{limited}~\cite{Ensign2018DecisionPrediction} or \textit{partial feedback}~\cite{Bechavod2019EqualFeedback} and the reason is that ML models are retrained using newly available data.
ML model feedback loops describe the case when the data that becomes newly available over time depends on the decision taken.
For example, hiring algorithms only learn about the skills of the candidates who were hired, credit lending algorithms only receive repaying probability information from people who received the loan, and predictive policing algorithms only register crime in patrolled neighborhoods.
In all these scenarios, the decision will create a gate to the pair $(x,y)$, which will be added to the existing data set $\left(X,Y\right)$ only when the decision is positive ($d=1$).
Notice that, when the retraining of the model does not depend on the decision (i.e., if the feature-label pair $(x,y)$ is added to the existing data set independently of $d$), there is no ML model feedback loop.
Using the language of dynamical systems theory, this case is simply viewed as an open-loop system with memory where the state variable $(X,Y)$ evolves according to the inner dynamics, but independently of the output variable (the decision $d$).
Additional examples of the ML model feedback loop can also be found in \cite{NIPS2016_962e56a8,Ensign2018RunawayPolicing,Elzayn2019FairProblems}.

\subsubsection{Outcome Feedback Loop}
\label{sssec:Outcome_Feedback_Loop}
Finally, in the outcome feedback loop, the decision $d$ affects the outcome $y$ before it is realized and ultimately observed.
Notice that this observed outcome then needs to be reused in some form in order to close the loop.
Namely, it only forms a loop if the outcome is used, e.g., as part of the training or validation data when retraining the model\footnote{Notice that this can, but does not need to, happen through an ML model feedback loop.}.
To see how an outcome feedback loop can arise, consider again the credit lending scenario: if a person is predicted at high risk of default, the loan might be granted, but at a higher interest rate.
However, the decision to enforce a higher interest rate further increases the chances that the customer defaults~\cite{Perdomo2020PerformativePrediction}.
In contrast to the example provided in Section~\ref{sssec:Feature_Feedback_Loop}, here we assume that the lender's decision $d$ has an effect on the realization of the outcome $y$, i.e., whether the loan is paid back or not, rather than on the features (credit score).

\subsection{Adversarial Feedback Loops}\label{ssec:adversarial_feedback_loops}

Some of the previously described feedback loops can take the form of what we call \textit{adversarial feedback loops}, which can arise if the decision $d$ is intertwined with an adversarial reaction to it.
While there is no visual difference with respect to the block-diagram representation, \textit{adversarial feedback loops} differ from their \textit{non-adversarial} counterpart in that the decision triggers the reaction of the individual(s) subjected to the decision-making process, which then affects the ML-pipeline.
In practice, in \textit{adversarial feedback loops}, individuals subjected to the decision-making process react strategically to the previous decisions by taking actions that increase their chances of receiving favorable decisions.
Notice that this is an important distinction for the design of measures that intend to control the system's dynamics, e.g., through bias mitigation techniques (as we will discuss in Section~\ref{sec:related_work} in more detail).
For instance, consider the attention allocation problem discussed in~\cite{DAmour2020FairnessStudies}.
Here, the decision-maker has limited (insufficient) resources to exhaustively inspect $N$ different locations, and therefore they have to decide where to (dynamically) allocate the attention.
As the authors argue, the incident rate of each of the $N$ sites responds dynamically (and adversarially) to the previous allocation, i.e., it increases where there was absolutely no control, and vice-versa it decreases proportionally to the amount of inspection.
In essence, this example describes the case of an \textit{adversarial individual feedback loop}, because the decision ultimately affects the incident rate, i.e., $\theta$. 

To give another example, consider a college that publishes the decision rule for its admission policy. 
Prospective students can strategically invest in their own qualifications in order to meet the requirements. 
If this action truly changes the preparation level of the student~\cite{Liu2020thedisparate}, then it is again an \textit{adversarial individual feedback loop}. 
However, it is also possible that only the observable features of the individual are changed~\cite{Heidari2019OnLearning}, e.g., if the students invest in SAT exam preparation without changing their actual qualification for the college.
Then, we are facing an \textit{adversarial feature feedback loop}.
Similarly, if an individual is applying for a loan, it might be beneficial to open multiple credit lines to improve their observable features~\cite{Perdomo2020PerformativePrediction}.
This action is not truly modifying the individual's capability of paying back the loan, but it is only a way to game the decision-making policy, thus we have an \textit{adversarial feature feedback loop}.

Additional examples of adversarial individual and feature feedback loops can be found in \cite{Hu2018AMarket,Heidari2019OnLearning,Kleinberg2020HowStrategically, Zhang2020HowQualification} and \cite{Hu2018TheManipulation,Tsirtsis2019OptimalBehavior,Milli2019,Kleinberg2020HowStrategically,DAmour2020FairnessStudies}, respectively.
However, we emphasize that it is not always easy to distinguish between the individual and the feature adversarial feedback loops, because many of these works 
assume that the decision affects the qualification $\theta$ of the individuals, but oftentimes they intend that it only affects its observable features $x$.

\subsection{Coexistence of Feedback Loops}
As seen in the previous sections, different feedback loops can coexist within the same application domain.
For instance, the recommender systems for an online platform can affect the opinion of the users $\theta$ (individual feedback loop) or just their representation in the feature space $x$ (feature feedback loop). College admission policies can induce students to improve their qualification (adversarial individual feedback loop) or just their representation $x$ (adversarial feature feedback loop). Alternatively, they can also lead to different retention rates across groups (sampling feedback loop).
Lending decisions can affect an individual's creditworthiness $\theta$ (individual feedback loop), credit score $x$ (feature feedback loop), realized outcome $y$ (i.e., whether or not the granted loan is paid back, representing an outcome feedback loop), or even the data used for the ML model development $(X,Y)$ (resulting in an ML model feedback loop) or the sample of individuals applying for a loan in the first place (causing a sampling feedback loop).
All five classified feedback loops represent some causal effect of the final decision on another component of the ML-based decision-making pipeline.
Thus, which type(s) of feedback loop(s) (co)exists solely depends on the context-specific assumptions regarding the underlying causal effects of the decision.
The possibility of the coexistence of different combinations of feedback loops gives rise to coupled behavior and even more complex dynamics.

\subsection{Positive/Negative Feedback Loops and Relation to Stability}
\label{ssec:positive_negative_fl}

In many disciplines, including the ML community, a considerable emphasis is placed on classifying feedback loops as either \emph{positive} or \emph{negative}~\cite{AR:83,NO-RM-JL-RPE-PK:11,JDS:00}.
This is often accompanied by some ambiguity in the definition of these notions.
In systems theory, a positive feedback loop (also known as \emph{reinforcing}) amplifies the effect of inputs on the outputs, while a negative feedback loop (also known as \emph{balancing}) attenuates it.
In other domains, the notion of a positive/negative feedback loop is sometimes associated with desirable/undesirable outcomes, regardless of how it acts to amplify/attenuate inputs.
For example, the feedback loop that increases recidivism due to incarcerated individuals' reduced access to finance is referred to as a negative feedback loop in~\cite[p.~2]{Zhang2021FairnessSurvey}.
This ambiguity is problematic, especially considering that in systems theory
the desired goal is often to make the output a predictable function of the input and independent from other exogenous but inevitable inputs (considered as \emph{disturbances}).
For this reason, properly designed negative feedback loops are deemed preferable,
while positive feedback loops are often considered problematic.

However, systems theory often places more emphasis on the \emph{stability} of the closed-loop system rather than classifying feedback loops as positive or negative.
A stable system converges to a predictable equilibrium point, while an unstable system either oscillates or grows beyond bounds.
It is intuitive to associate positive feedback with instability and negative feedback with stability, however, this intuition is not universal~\cite{aastrom2021feedback,zeigler2000theory}.
On the one hand, positive feedback is guaranteed to lead to instability only in the special class of linear systems.
The presence of non-linearity (e.g., saturation or hysteresis) can stabilize a positive feedback loop, which is intentionally introduced in many cases (e.g., the design of signal amplifiers).
On the other hand, negative feedback does not guarantee stability (even in linear systems).
Moreover, the same system could be in either positive or negative feedback depending on the operating regime (e.g., the frequency of the input signal).
Thus, in this paper, we shift the focus from classifying feedback loops as positive/negative to asking whether the closed-loop system converges (or not) to a (desirable) state.
As we will see in the examples in Section~\ref{sec:example}, feedback loops often drive the ML-based decision system to stable equilibrium points in the long run.

\section{Feedback Loops and Algorithmic Biases}
\label{sec:implications-algo-fairness}

Being able to reason about what caused certain types of bias is of incredible practical importance in order to avoid or counteract them in the long term.
Otherwise, ML-based decision-making systems can result in socially undesirable outcomes over time.
Many works claim that those biases can be perpetuated or even reinforced due to feedback loops~\cite{Lum2016Topredict,oneil2017weapons,barocas-hardt-narayanan,Chouldechova2018,Chouldechova2020,mehrabi2021survey,Kearns2019EthicalAlgorithm,olteanu2019socialdata,VanGiffen2022biases}.
However, a clear understanding of the causal effects of feedback loops on algorithmic biases is currently missing.
We fill this gap by connecting the classification of feedback loops (which we introduced in Section~\ref{ssec:Feedback-loops}) to algorithmic biases and explain in more detail \textit{which} types of bias they affect.
Table~\ref{tab:feedback-loops-and-ML-biases} provides an overview of the connections we establish.
However, the term `bias' can have different meanings and be used interchangeably with synonyms for different types of bias. To ensure consistency, we adopt the concepts and terminology introduced by~\citet{10.1145/3465416.3483305}.

\begin{table*}[ht]
\centering
\caption{Feedback loops and the ML biases they affect}
\label{tab:feedback-loops-and-ML-biases}
\begin{tabular}{ll}
\toprule
\textbf{Feedback loop}   & \textbf{ML bias}                           \\
\midrule
Sampling, ML model   & {Representation bias}                            \\  \hline
Individual & Historical bias                                                  \\ \hline
Feature, Outcome    & {Measurement bias}           \\
\bottomrule
\end{tabular}
\end{table*}

\paragraph{Representation Bias}
    According to~\cite{10.1145/3465416.3483305}, there are different nuances of representation bias: Representation bias can arise (i) if the defined target population does not reflect the use population, (ii) if the target population contains underrepresented groups, and (iii) if the sampled group of individuals is not representative of the target population.
    All three versions represent some difference between the used dataset $(X,Y)$ and the population $\mathcal{I}$.

    \textit{Sampling feedback loops} can affect representation bias.
    Sampling feedback loops affect the sampling function $s$ that outputs a set of individuals on which an ML-based decision-making system acts.
    A sampling feedback loop changes the sample of individuals for whom a prediction and, ultimately, a decision is made (i.e., those who get a chance to be selected).
    Thus, it can result in representation bias, which describes the situation in which $s$ undersamples some part of the population.
    As a result, the available data is not representative of $\mathcal{I}$ and, for this reason, the ML model likely does not generalize well for the disadvantaged group~\cite{10.1145/3465416.3483305}.
    
    \textit{ML model feedback loops} can also affect representation bias.
    The ML model feedback loop changes the sample of individuals whose realized outcome becomes observable, i.e., those that are selected and can thus be added as a new feature-label pair $(x,y)$ to the sample $(X,Y)$ -- see Fig.~\ref{fig:pipeline_MLmodel} in Appendix~\ref{app:detailed_ML_pipeline} for a visualization of this process.
    Therefore, it can affect representation bias, which stems from a shift in the training data distributions.%
    \footnote{
    Additionally, ML model feedback loops can affect evaluation bias.
    More generally, evaluation bias exists if the sample used to evaluate an ML-based decision system does not represent the population it is used for, i.e., it stems from a shift in the evaluation data distributions~\cite{olteanu2019socialdata,10.1145/3465416.3483305,mehrabi2021survey,VanGiffen2022biases}.
    Hence, ML model feedback loops can affect evaluation bias if the sample used for evaluation $(X_m,Y_m)$ is influenced by past decisions.
    }

\paragraph{Historical Bias}
    \textit{Individual feedback loops} can affect historical bias.
    \textit{Individual feedback loops} act on the \textit{construct space} (CS) of an individual, i.e., the inherent properties of an individual $\theta$ change and not only the observed proxies $x,y$, which are measured in the \textit{observed space} (OS)~\cite{Friedler2016}.
    This can result in historical bias (also called ``life bias''~\cite{Hertweck2021statistical-parity}), which describes injustices that manifest in inequality between groups in the CS.
    As decisions can change individuals' properties $\theta$, which can manifest in altered future features $x$, it becomes more difficult to treat individuals fairly since the decision actually changed them.
    This means that the world is accurately represented by the data (i.e., the measurement functions $r$ and $t$ are acceptable), but the state of the world (i.e., an individual's inherent decision-relevant attributes $\theta$) is the result of unfair treatments in previous decision rounds~\cite{10.1145/3465416.3483305}.
    For example, not considering counterfactual decisions for individuals (i.e., assuming that individuals would have evolved identically over time, even if they had been assigned different decisions) can drive the decision system to a state in which individuals are disadvantaged solely because of an unlucky event in the past, even if their attributes are perfectly measurable.

\paragraph{Measurement Bias}
    \textit{Outcome feedback loops} and \textit{feature feedback loops} can affect measurement bias~\cite{Friedler2016,olteanu2019socialdata,10.1145/3465416.3483305,mehrabi2021survey,VanGiffen2022biases}.
    These two feedback loops act on the measurement functions $r$ and $t$ and thus affect an individual's observable properties $x,a,y$.
    The outcome feedback loop changes the realization of the outcome ($y$).
    In contrast, the feature feedback loop changes the observable attributes that are fed into the prediction model ($x$ and, potentially, $a$), i.e., the features for future decisions.
    Thus, both types of feedback loops can affect measurement bias:
    the features $x$ and labels $y$ are usually just proxies as they try to measure an inherent property of an individual, which might represent a construct that is not directly measurable or even observable ($\theta$)~\cite{10.1145/3465416.3483305}.
    Measurement bias describes the transition between CS and OS~\cite{Friedler2016}.
    Thus, it describes a situation in which those proxies less closely approximate the intended attribute for certain individuals or groups, which means that $r$ or $t$ (or both) are not appropriate to capture the relevant construct.%
    \footnote{
    Notice that the list of the biases in Table~\ref{tab:feedback-loops-and-ML-biases} is non-exhaustive since some biases are connected: feedback loops could also indirectly affect learning bias or aggregation bias.
    Learning bias represents a limitation of the learned function $f$ that occurs when erroneously assuming that $p(y|x)$ is homogeneous across groups~\cite{10.1145/3465416.3483305}.
    Aggregation bias arises if the ML-based system fails to draw the correct conclusions for certain individuals and, therefore, results in disproportionately worse decisions for some group~\cite{10.1145/3465416.3483305}.
    For example, feature or outcome feedback loops directly affect measurement bias and indirectly affect learning bias at the same time if the shift of the distribution $(X,Y)$ -- which is connected to measurement bias -- also results in heterogeneous conditional probabilities, $p(y|x)$, of getting a certain output for a given input across groups.
    Similarly, they could indirectly result in learning bias if the way the learned function $f$ is optimized is less suited for the new distribution $(X,Y)$.
    }
    For example, using arrests as a proxy for the risk of committing a crime (as is the case in the recidivism risk prediction tool COMPAS~\cite{angwin2016machine}) is problematic if there are groups that are much more likely to be arrested for certain crimes.

\section{Case Study: Feedback Loops in Recommender Systems}\label{sec:example}

To demonstrate the potential of our classification of feedback loops and their relation to the different biases, we present a unifying case study on recommender systems (RS).%
\footnote{
The code to simulate this case study is publicly available at \url{https://github.com/paganick/feedback-loops-and-biases}.
}
We consider the case of an online platform where the RS is used to provide content the users are interested in. For simplicity, we consider just one relevant item (e.g., a specific video) and denote a user's interest in this item with $\theta \in [0,1]$, where larger $\theta$ corresponds to higher interest.
The realized outcome $y$ denotes whether a user shows interest (e.g., clicks on the relevant item in question), $y=1$, or not, $y=0$.
The platform uses an RS to predict a user's interest $\hat{y}=f(x)$, where the feature $x \in [0,1]$ represents the user's past clicking behavior on the platform.
For this simple example, $x$ is the percentage of recommended relevant items that the user has clicked on in the past and thus serves as a proxy of the user's interest in the relevant item.
The function $f:[0,1] \rightarrow [0,1]$ is learned through a logistic regression (LR) algorithm (which is fitted to a sigmoid function) trained on data $(X,Y)$, which consists of a collection of feature-label pairs $\left(x,y\right)$.
To decide whether the relevant item should be shown as one of the top recommendations ($d=1$) or not ($d=0$), the following threshold rule is used: $d=1$ if $\hat{y} > 0.5$, and $d=0$ otherwise.
After each recommendation round, $y$ is observed, $(x,y)$ is added to the existing dataset $(X,Y)$, and the RS is retrained.
We consider two groups of users $a \in \{\text{G}1, \text{G}2\}$.
For simplicity, $a$ is not used as an input for the RS.

We now provide one example for each type of feedback loop described in Section~\ref{ssec:Feedback-loops} to illustrate how they are associated with different biases.
The initial conditions specific to each of these simulation examples are described in Table~\ref{tab:case_study_initial_conditions} and the initial $\theta$ distribution is shown in Fig.~\ref{fig:initial_theta_distribution} in Appendix~\ref{app:case_study_addendum}.
Notice that the mean of $\theta$ is higher for group G1 (i.e., $\mu_{\theta,G1}=0.7, \mu_{\theta,G2}=0.3$), which means that individuals of group G1 are more interested in the item, on average.

\begin{table*}[t]
\centering
\caption{Initial conditions for the different experiments. The acronyms stand for group 1 (G1), group 2 (G2), population size ($n$), training sample size ($n_\text{train}$), distribution mean ($\mu$) and standard deviation ($\sigma$), distribution of the feature realization ($r$), distribution of the outcome realization during the simulation ($t$) and for the initial training set ($t_\text{train}$).
For all experiments, we set the following parameters:
$n=1000$,
$n_{\text{train},G1} = n_{\text{train},G2}=500$,
$\sigma_{\theta,G2}=0.15$,
$\sigma_{t,G1}=0.1$,
$\mu_{r,G1}=0$,
$\mu_{t,G1}=0$,
$\sigma_{\theta,G1}=0.15$,
$\mu_{t,G2}=0$,
$\sigma_{t,G2}=0.1$,
$\mu_{t_\text{train}}=0$.
In the table, we describe the parameters that vary from one experiment to another.
}
\begin{tabular}{lcccccc}
\toprule
\textbf{\textbf{Feedback loop}} &
  \multicolumn{1}{r}{\textbf{$\mu_{\theta,G1}$}} &
  \multicolumn{1}{r}{$\mu_{\theta,G2}$} &
  \multicolumn{1}{r}{$\sigma_{r,G1}$} &
  \multicolumn{1}{r}{$\mu_{r,G2}$} &
  \multicolumn{1}{r}{$\sigma_{r,G2}$} &
  \multicolumn{1}{r}{$\sigma_{t_\text{train}}$} \\
\midrule
Sampling, Individual, Outcome
&
  \multirow{2}{*}{0.7} &
  \multirow{2}{*}{0.3} &
  \multirow{2}{*}{0.0} &
  \multirow{2}{*}{0.0} &
  \multirow{2}{*}{0.0} &
  0 \\ \cline{1-1} \cline{7-7}
ML model 
&     &     &     &      &     & 1 \\ \hline
Feature 
& 0.5 & 0.5 & 0.1 & -0.2 & 0.1 & 0 \\
\bottomrule
\end{tabular}
\label{tab:case_study_initial_conditions}
\end{table*}

\begin{figure}[!h]
\centering
\begin{subfigure}[t]{0.44\textwidth}
    \centering
\begin{tikzpicture}

\definecolor{color0}{rgb}{0.843137254901961,0.0980392156862745,0.109803921568627}
\definecolor{color1}{rgb}{0.172549019607843,0.482352941176471,0.713725490196078}

\tikzstyle{every node}=[font=\small]

\begin{axis}[
height=\plotheight,
width=\plotwidth,
legend cell align={left},
legend style={fill opacity=0.8, draw opacity=1, text opacity=1, draw=white!80!black},
legend pos=north west,
tick align=outside,
tick pos=left,
x grid style={white!69.0196078431373!black},
xlabel={time-steps},
xmin=-0.64, xmax=4.64,
xtick style={color=black},
xtick={0,1,2,3,4},
xticklabels={0,2000,10000,20000,50000},
y grid style={white!69.0196078431373!black},
ylabel={number of users},
ymin=0, ymax=999,
ytick style={color=black}
]
\draw[draw=none,fill=color0] (axis cs:-0.4,0) rectangle (axis cs:0,496);
\addlegendimage{ybar,ybar legend,draw=none,fill=color0};
\addlegendentry{Group 1}

\draw[draw=none,fill=color0] (axis cs:0.6,0) rectangle (axis cs:1,760);
\draw[draw=none,fill=color0] (axis cs:1.6,0) rectangle (axis cs:2,911);
\draw[draw=none,fill=color0] (axis cs:2.6,0) rectangle (axis cs:3,913);
\draw[draw=none,fill=color0] (axis cs:3.6,0) rectangle (axis cs:4,914);
\draw[draw=none,fill=color1] (axis cs:-2.77555756156289e-17,0) rectangle (axis cs:0.4,504);
\addlegendimage{ybar,ybar legend,draw=none,fill=color1};
\addlegendentry{Group 2}

\draw[draw=none,fill=color1] (axis cs:1,0) rectangle (axis cs:1.4,240);
\draw[draw=none,fill=color1] (axis cs:2,0) rectangle (axis cs:2.4,89);
\draw[draw=none,fill=color1] (axis cs:3,0) rectangle (axis cs:3.4,87);
\draw[draw=none,fill=color1] (axis cs:4,0) rectangle (axis cs:4.4,86);
\end{axis}

\end{tikzpicture}
    \caption{\textbf{Sampling FL}: platform user cardinalities}
    \label{fig:sampling_FL_users}
\end{subfigure}%
\hfill{}
\begin{subfigure}[t]{0.44\textwidth}
    \centering
\begin{tikzpicture}

\definecolor{color0}{rgb}{0.843137254901961,0.0980392156862745,0.109803921568627}
\definecolor{color1}{rgb}{0.172549019607843,0.482352941176471,0.713725490196078}

\tikzstyle{every node}=[font=\small]

\begin{axis}[
height=\plotheight,
width=\plotwidth,
legend cell align={left},
legend style={fill opacity=0.8, draw opacity=1, text opacity=1, at={(0.97,0.03)}, anchor=south east, draw=white!80!black},
tick align=outside,
tick pos=left,
x grid style={white!69.0196078431373!black},
xlabel={time-steps},
xmin=-0.9, xmax=8.9,
xtick style={color=black},
xtick={0,2,4,6,8},
xticklabels={0,2000,10000,20000,50000},
y grid style={white!69.0196078431373!black},
ylabel={$\theta$},
ymin=-0.006243454720707, ymax=1.04525015022686,
ytick style={color=black},
ytick={-0.2,0,0.2,0.4,0.6,0.8,1,1.2},
yticklabels={−0.2,0.0,0.2,0.4,0.6,0.8,1.0,1.2},
]
\addplot [color0, forget plot]
table {%
-0.7 0.598016010658195
-0.1 0.598016010658195
-0.1 0.796960203024725
-0.7 0.796960203024725
-0.7 0.598016010658195
};
\label{p1}
\addplot [color0, forget plot]
table {%
-0.4 0.598016010658195
-0.4 0.344258582175875
};
\addplot [color0, forget plot]
table {%
-0.4 0.796960203024725
-0.4 0.998064856330019
};
\addplot [color0, forget plot]
table {%
-0.55 0.344258582175875
-0.25 0.344258582175875
};
\addplot [color0, forget plot]
table {%
-0.55 0.998064856330019
-0.25 0.998064856330019
};
\addplot [color0, mark=*, mark size=1, mark options={solid,fill opacity=0}, only marks, forget plot]
table {%
-0.4 0.250118627285822
-0.4 0.209289347920417
-0.4 0.192781920195074
};
\addplot [color0, forget plot]
table {%
1.3 0.620198017252052
1.9 0.620198017252052
1.9 0.809435317393545
1.3 0.809435317393545
1.3 0.620198017252052
};
\addplot [color0, forget plot]
table {%
1.6 0.620198017252052
1.6 0.356952053496971
};
\addplot [color0, forget plot]
table {%
1.6 0.809435317393545
1.6 0.998064856330019
};
\addplot [color0, forget plot]
table {%
1.45 0.356952053496971
1.75 0.356952053496971
};
\addplot [color0, forget plot]
table {%
1.45 0.998064856330019
1.75 0.998064856330019
};
\addplot [color0, mark=*, mark size=1, mark options={solid,fill opacity=0}, only marks, forget plot]
table {%
1.6 0.317278686415049
1.6 0.250118627285822
1.6 0.209289347920417
};
\addplot [color0, forget plot]
table {%
3.3 0.625381456808518
3.9 0.625381456808518
3.9 0.805784513737897
3.3 0.805784513737897
3.3 0.625381456808518
};
\addplot [color0, forget plot]
table {%
3.6 0.625381456808518
3.6 0.512900577220146
};
\addplot [color0, forget plot]
table {%
3.6 0.805784513737897
3.6 0.998064856330019
};
\addplot [color0, forget plot]
table {%
3.45 0.512900577220146
3.75 0.512900577220146
};
\addplot [color0, forget plot]
table {%
3.45 0.998064856330019
3.75 0.998064856330019
};
\addplot [color0, forget plot]
table {%
5.3 0.626045076843442
5.9 0.626045076843442
5.9 0.805884533695905
5.3 0.805884533695905
5.3 0.626045076843442
};
\addplot [color0, forget plot]
table {%
5.6 0.626045076843442
5.6 0.51398450187718
};
\addplot [color0, forget plot]
table {%
5.6 0.805884533695905
5.6 0.998064856330019
};
\addplot [color0, forget plot]
table {%
5.45 0.51398450187718
5.75 0.51398450187718
};
\addplot [color0, forget plot]
table {%
5.45 0.998064856330019
5.75 0.998064856330019
};
\addplot [color0, forget plot]
table {%
7.3 0.626060160757259
7.9 0.626060160757259
7.9 0.805834523716901
7.3 0.805834523716901
7.3 0.626060160757259
};
\addplot [color0, forget plot]
table {%
7.6 0.626060160757259
7.6 0.51398450187718
};
\addplot [color0, forget plot]
table {%
7.6 0.805834523716901
7.6 0.998064856330019
};
\addplot [color0, forget plot]
table {%
7.45 0.51398450187718
7.75 0.51398450187718
};
\addplot [color0, forget plot]
table {%
7.45 0.998064856330019
7.75 0.998064856330019
};
\addplot [color1, forget plot]
table {%
0.1 0.204613745038969
0.7 0.204613745038969
0.7 0.418222143702826
0.1 0.418222143702826
0.1 0.204613745038969
};
\addplot [color1, forget plot]
table {%
0.4 0.204613745038969
0.4 0.0117855441296537
};
\addplot [color1, forget plot]
table {%
0.4 0.418222143702826
0.4 0.737695241335512
};
\addplot [color1, forget plot]
table {%
0.25 0.0117855441296537
0.55 0.0117855441296537
};
\addplot [color1, forget plot]
table {%
0.25 0.737695241335512
0.55 0.737695241335512
};
\addplot [color1, mark=*, mark size=1, mark options={solid,fill opacity=0}, only marks, forget plot]
table {%
0.4 0.758130653471923
0.4 0.741909737030158
};
\addplot [color1, forget plot]
table {%
2.1 0.254519816939162
2.7 0.254519816939162
2.7 0.532476798414333
2.1 0.532476798414333
2.1 0.254519816939162
};
\addplot [color1, forget plot]
table {%
2.4 0.254519816939162
2.4 0.0116086502466546
};
\addplot [color1, forget plot]
table {%
2.4 0.532476798414333
2.4 0.770152339649816
};
\addplot [color1, forget plot]
table {%
2.25 0.0116086502466546
2.55 0.0116086502466546
};
\addplot [color1, forget plot]
table {%
2.25 0.770152339649816
2.55 0.770152339649816
};
\addplot [color1, forget plot]
table {%
4.1 0.538241876045127
4.7 0.538241876045127
4.7 0.628383506869582
4.1 0.628383506869582
4.1 0.538241876045127
};
\addplot [color1, forget plot]
table {%
4.4 0.538241876045127
4.4 0.513055752482519
};
\addplot [color1, forget plot]
table {%
4.4 0.628383506869582
4.4 0.758436795774358
};
\addplot [color1, forget plot]
table {%
4.25 0.513055752482519
4.55 0.513055752482519
};
\addplot [color1, forget plot]
table {%
4.25 0.758436795774358
4.55 0.758436795774358
};
\addplot [color1, mark=*, mark size=1, mark options={solid,fill opacity=0}, only marks, forget plot]
table {%
4.4 0.375229343164101
4.4 0.392773371165612
4.4 0.304929077178937
4.4 0.770152339649816
};
\addplot [color1, forget plot]
table {%
6.1 0.538802814604917
6.7 0.538802814604917
6.7 0.628612002321103
6.1 0.628612002321103
6.1 0.538802814604917
};
\addplot [color1, forget plot]
table {%
6.4 0.538802814604917
6.4 0.513055752482519
};
\addplot [color1, forget plot]
table {%
6.4 0.628612002321103
6.4 0.758436795774358
};
\addplot [color1, forget plot]
table {%
6.25 0.513055752482519
6.55 0.513055752482519
};
\addplot [color1, forget plot]
table {%
6.25 0.758436795774358
6.55 0.758436795774358
};
\addplot [color1, mark=*, mark size=1, mark options={solid,fill opacity=0}, only marks, forget plot]
table {%
6.4 0.304929077178937
6.4 0.770152339649816
};
\addplot [color1, forget plot]
table {%
8.1 0.539423517681506
8.7 0.539423517681506
8.7 0.628726250046864
8.1 0.628726250046864
8.1 0.539423517681506
};
\addplot [color1, forget plot]
table {%
8.4 0.539423517681506
8.4 0.514994185756079
};
\addplot [color1, forget plot]
table {%
8.4 0.628726250046864
8.4 0.758436795774358
};
\addplot [color1, forget plot]
table {%
8.25 0.514994185756079
8.55 0.514994185756079
};
\addplot [color1, forget plot]
table {%
8.25 0.758436795774358
8.55 0.758436795774358
};
\addplot [color1, mark=*, mark size=1, mark options={solid,fill opacity=0}, only marks, forget plot]
table {%
8.4 0.304929077178937
8.4 0.770152339649816
};
\addplot [color0, forget plot]
table {%
-0.7 0.688024954149869
-0.1 0.688024954149869
};
\addplot [color0, forget plot]
table {%
1.3 0.706866563039477
1.9 0.706866563039477
};
\addplot [color0, forget plot]
table {%
3.3 0.713386197022475
3.9 0.713386197022475
};
\addplot [color0, forget plot]
table {%
5.3 0.713386197022475
5.9 0.713386197022475
};
\addplot [color0, forget plot]
table {%
7.3 0.712858062580034
7.9 0.712858062580034
};
\addplot [color1, forget plot]
table {%
0.1 0.310561102488851
0.7 0.310561102488851
};
\addplot [color1, forget plot]
table {%
2.1 0.396178188784364
2.7 0.396178188784364
};
\addplot [color1, forget plot]
table {%
4.1 0.565109426452076
4.7 0.565109426452076
};
\addplot [color1, forget plot]
table {%
6.1 0.565538080947363
6.7 0.565538080947363
};
\addplot [color1, forget plot]
table {%
8.1 0.565598149365034
8.7 0.565598149365034
};
\label{p2}
\addplot [semithick, black, dashed, forget plot]
table {%
-0.9 0.5
8.9 0.5
};
\end{axis}

\node [
fill opacity=0.8, draw opacity=1, text opacity=1, draw=white!80!black] at (rel axis cs: 0.7,0.14) {\shortstack[l]{
\ref{p1} Group 1 \hspace{3pt}
\ref{p2} Group 2 
}};

\end{tikzpicture}
    \caption{\textbf{Sampling FL}: interests of platform users}
    \label{fig:sampling_FL_thetas}
\end{subfigure}%
\vspace{10pt}
\begin{subfigure}[t]{0.44\textwidth}
    \centering
    \input{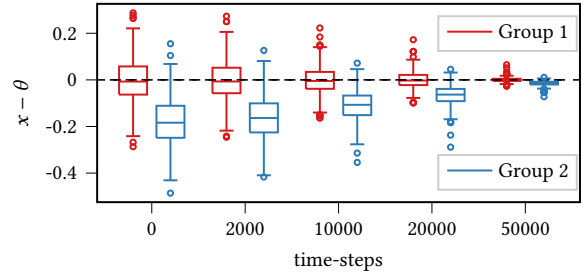}
    \caption{\textbf{Individual FL}: interests of platform users}
    \label{fig:individual_FL_thetas}
\end{subfigure}%
\hfill{}
\begin{subfigure}[t]{0.44\textwidth}
    \centering
    \input{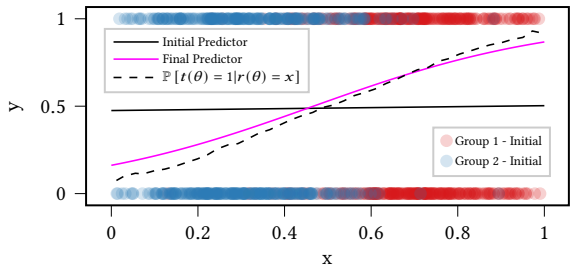}
    \caption{\textbf{Feature FL}: measurement error $\left( x - \theta \right)$}
    \label{fig:feature_FL_thetas}
\end{subfigure}%
\vspace{10pt}
\begin{subfigure}[t]{0.44\textwidth}
    \centering
    \input{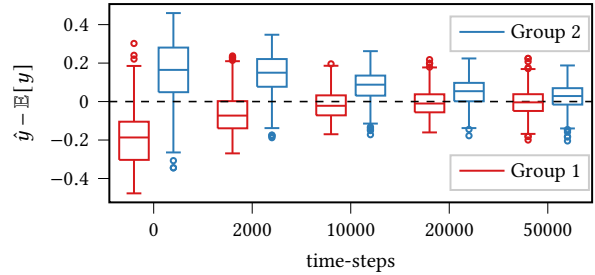}
    \caption{\textbf{ML model FL}: initial distribution of $(X,Y)$, initial/final predictors, and outcome realization $t$}
    \label{fig:ML_FL_predictions}
\end{subfigure}%
\hfill{}
\begin{subfigure}[t]{0.44\textwidth}
    \centering
\begin{tikzpicture}

\definecolor{color0}{rgb}{0.843137254901961,0.0980392156862745,0.109803921568627}
\definecolor{color1}{rgb}{0.172549019607843,0.482352941176471,0.713725490196078}

\tikzstyle{every node}=[font=\small]

\begin{axis}[
height=\plotheight,
width=\plotwidth,
legend cell align={left},
legend style={fill opacity=0.8, draw opacity=1, text opacity=1, draw=white!80!black},
tick align=outside,
tick pos=left,
x grid style={white!69.0196078431373!black},
xlabel={time-steps},
xmin=-0.9, xmax=8.9,
xtick style={color=black},
xtick={0,2,4,6,8},
xticklabels={0,2000,10000,20000,50000},
y grid style={white!69.0196078431373!black},
ylabel={\(\displaystyle \hat{y}-\mathbb{E}[y]\)},
ymin=-0.524096977229325, ymax=0.506862785413115,
ytick style={color=black},
ytick={-0.6,-0.4,-0.2,0,0.2,0.4,0.6},
]
\addplot [color0, forget plot]
table {%
-0.7 -0.303177791842044
-0.1 -0.303177791842044
-0.1 -0.104682902601551
-0.7 -0.104682902601551
-0.7 -0.303177791842044
};
\label{p1}
\addplot [color0, forget plot]
table {%
-0.4 -0.303177791842044
-0.4 -0.477235169836487
};
\addplot [color0, forget plot]
table {%
-0.4 -0.104682902601551
-0.4 0.185123318015275
};
\addplot [color0, forget plot]
table {%
-0.55 -0.477235169836487
-0.25 -0.477235169836487
};
\addplot [color0, forget plot]
table {%
-0.55 0.185123318015275
-0.25 0.185123318015275
};
\addplot [black, mark=*, mark size=1, mark options={solid,fill opacity=0,draw=color0}, only marks, forget plot]
table {%
-0.4 0.302062151380567
-0.4 0.220927988138257
-0.4 0.240469496900834
};
\addplot [color0, forget plot]
table {%
1.3 -0.139044132239014
1.9 -0.139044132239014
1.9 0.00224077631862787
1.3 0.00224077631862787
1.3 -0.139044132239014
};
\addplot [color0, forget plot]
table {%
1.6 -0.139044132239014
1.6 -0.269372784615538
};
\addplot [color0, forget plot]
table {%
1.6 0.00224077631862787
1.6 0.209865841474039
};
\addplot [color0, forget plot]
table {%
1.45 -0.269372784615538
1.75 -0.269372784615538
};
\addplot [color0, forget plot]
table {%
1.45 0.209865841474039
1.75 0.209865841474039
};
\addplot [black, mark=*, mark size=1, mark options={solid,fill opacity=0,draw=color0}, only marks, forget plot]
table {%
1.6 0.228774350636625
1.6 0.21443740432884
1.6 0.236899713709668
1.6 0.230351439348486
1.6 0.220998260678999
};
\addplot [color0, forget plot]
table {%
3.3 -0.0717034081221682
3.9 -0.0717034081221682
3.9 0.0322554484189832
3.3 0.0322554484189832
3.3 -0.0717034081221682
};
\addplot [color0, forget plot]
table {%
3.6 -0.0717034081221682
3.6 -0.169880031220905
};
\addplot [color0, forget plot]
table {%
3.6 0.0322554484189832
3.6 0.186075455331357
};
\addplot [color0, forget plot]
table {%
3.45 -0.169880031220905
3.75 -0.169880031220905
};
\addplot [color0, forget plot]
table {%
3.45 0.186075455331357
3.75 0.186075455331357
};
\addplot [black, mark=*, mark size=1, mark options={solid,fill opacity=0,draw=color0}, only marks, forget plot]
table {%
3.6 0.195892076795818
};
\addplot [color0, forget plot]
table {%
5.3 -0.0559346613028585
5.9 -0.0559346613028585
5.9 0.0378219964904636
5.3 0.0378219964904636
5.3 -0.0559346613028585
};
\addplot [color0, forget plot]
table {%
5.6 -0.0559346613028585
5.6 -0.160428507896586
};
\addplot [color0, forget plot]
table {%
5.6 0.0378219964904636
5.6 0.177365175890121
};
\addplot [color0, forget plot]
table {%
5.45 -0.160428507896586
5.75 -0.160428507896586
};
\addplot [color0, forget plot]
table {%
5.45 0.177365175890121
5.75 0.177365175890121
};
\addplot [black, mark=*, mark size=1, mark options={solid,fill opacity=0,draw=color0}, only marks, forget plot]
table {%
5.6 0.17922616434699
5.6 0.198383498008286
5.6 0.217904662616463
5.6 0.19929496933414
};
\addplot [color0, forget plot]
table {%
7.3 -0.0488831858630229
7.9 -0.0488831858630229
7.9 0.0382807272359953
7.3 0.0382807272359953
7.3 -0.0488831858630229
};
\addplot [color0, forget plot]
table {%
7.6 -0.0488831858630229
7.6 -0.168140586778429
};
\addplot [color0, forget plot]
table {%
7.6 0.0382807272359953
7.6 0.169005328716974
};
\addplot [color0, forget plot]
table {%
7.45 -0.168140586778429
7.75 -0.168140586778429
};
\addplot [color0, forget plot]
table {%
7.45 0.169005328716974
7.75 0.169005328716974
};
\addplot [black, mark=*, mark size=1, mark options={solid,fill opacity=0,draw=color0}, only marks, forget plot]
table {%
7.6 -0.181635381550578
7.6 -0.199902276164059
7.6 -0.18071382431824
7.6 0.224572335689686
7.6 0.207706441075797
7.6 0.171254776397958
7.6 0.176000695749954
7.6 0.223539623674142
};
\addplot [color1, forget plot]
table {%
0.1 0.0487062467998819
0.7 0.0487062467998819
0.7 0.280656661257027
0.1 0.280656661257027
0.1 0.0487062467998819
};
\addplot [color1, forget plot]
table {%
0.4 0.0487062467998819
0.4 -0.264523641844008
};
\addplot [color1, forget plot]
table {%
0.4 0.280656661257027
0.4 0.460000978020277
};
\addplot [color1, forget plot]
table {%
0.25 -0.264523641844008
0.55 -0.264523641844008
};
\addplot [color1, forget plot]
table {%
0.25 0.460000978020277
0.55 0.460000978020277
};
\addplot [black, mark=*, mark size=1, mark options={solid,fill opacity=0,draw=color1}, only marks, forget plot]
table {%
0.4 -0.307187913683346
0.4 -0.344504679875193
0.4 -0.344382862135545
};
\addplot [color1, forget plot]
table {%
2.1 0.0769466304558141
2.7 0.0769466304558141
2.7 0.221022118816809
2.1 0.221022118816809
2.1 0.0769466304558141
};
\addplot [color1, forget plot]
table {%
2.4 0.0769466304558141
2.4 -0.137990478985926
};
\addplot [color1, forget plot]
table {%
2.4 0.221022118816809
2.4 0.347402714260906
};
\addplot [color1, forget plot]
table {%
2.25 -0.137990478985926
2.55 -0.137990478985926
};
\addplot [color1, forget plot]
table {%
2.25 0.347402714260906
2.55 0.347402714260906
};
\addplot [black, mark=*, mark size=1, mark options={solid,fill opacity=0,draw=color1}, only marks, forget plot]
table {%
2.4 -0.187760156248165
2.4 -0.18090077609771
2.4 -0.174253042904647
2.4 -0.175948960607998
};
\addplot [color1, forget plot]
table {%
4.1 0.0307814743964782
4.7 0.0307814743964782
4.7 0.13510474117636
4.1 0.13510474117636
4.1 0.0307814743964782
};
\addplot [color1, forget plot]
table {%
4.4 0.0307814743964782
4.4 -0.114475598253757
};
\addplot [color1, forget plot]
table {%
4.4 0.13510474117636
4.4 0.262040367219349
};
\addplot [color1, forget plot]
table {%
4.25 -0.114475598253757
4.55 -0.114475598253757
};
\addplot [color1, forget plot]
table {%
4.25 0.262040367219349
4.55 0.262040367219349
};
\addplot [black, mark=*, mark size=1, mark options={solid,fill opacity=0,draw=color1}, only marks, forget plot]
table {%
4.4 -0.145824876988959
4.4 -0.137600147298275
4.4 -0.131832824099731
4.4 -0.172144451536607
4.4 -0.150739340507679
};
\addplot [color1, forget plot]
table {%
6.1 0.00147649484413813
6.7 0.00147649484413813
6.7 0.0971362369889619
6.1 0.0971362369889619
6.1 0.00147649484413813
};
\addplot [color1, forget plot]
table {%
6.4 0.00147649484413813
6.4 -0.137864421935778
};
\addplot [color1, forget plot]
table {%
6.4 0.0971362369889619
6.4 0.224119758794538
};
\addplot [color1, forget plot]
table {%
6.25 -0.137864421935778
6.55 -0.137864421935778
};
\addplot [color1, forget plot]
table {%
6.25 0.224119758794538
6.55 0.224119758794538
};
\addplot [black, mark=*, mark size=1, mark options={solid,fill opacity=0,draw=color1}, only marks, forget plot]
table {%
6.4 -0.177516682201782
6.4 -0.144710607174891
};
\addplot [color1, forget plot]
table {%
8.1 -0.0157412165429246
8.7 -0.0157412165429246
8.7 0.0696344495477822
8.1 0.0696344495477822
8.1 -0.0157412165429246
};
\addplot [color1, forget plot]
table {%
8.4 -0.0157412165429246
8.4 -0.139751732427276
};
\addplot [color1, forget plot]
table {%
8.4 0.0696344495477822
8.4 0.187712007323171
};
\addplot [color1, forget plot]
table {%
8.25 -0.139751732427276
8.55 -0.139751732427276
};
\addplot [color1, forget plot]
table {%
8.25 0.187712007323171
8.55 0.187712007323171
};
\addplot [black, mark=*, mark size=1, mark options={solid,fill opacity=0,draw=color1}, only marks, forget plot]
table {%
8.4 -0.146254590904667
8.4 -0.158600607793759
8.4 -0.204657195714695
8.4 -0.184173104165291
8.4 -0.146993208202797
8.4 -0.181965492115088
};
\addplot [semithick, black, dashed, forget plot]
table {%
-0.9 0
8.9 0
};
\addplot [color0, forget plot]
table {%
-0.7 -0.186676288460355
-0.1 -0.186676288460355
};
\addplot [color0, forget plot]
table {%
1.3 -0.0729457227326817
1.9 -0.0729457227326817
};
\addplot [color0, forget plot]
table {%
3.3 -0.0219930309669663
3.9 -0.0219930309669663
};
\addplot [color0, forget plot]
table {%
5.3 -0.0101535292422892
5.9 -0.0101535292422892
};
\addplot [color0, forget plot]
table {%
7.3 -0.00412253852080952
7.9 -0.00412253852080952
};
\addplot [color1, forget plot]
table {%
0.1 0.164660503024996
0.7 0.164660503024996
};
\addplot [color1, forget plot]
table {%
2.1 0.150010297944747
2.7 0.150010297944747
};
\addplot [color1, forget plot]
table {%
4.1 0.0877879205878141
4.7 0.0877879205878141
};
\addplot [color1, forget plot]
table {%
6.1 0.0539278827879523
6.7 0.0539278827879523
};
\addplot [color1, forget plot]
table {%
8.1 0.0286830628965035
8.7 0.0286830628965035
};
\label{p2}
\end{axis}

\node [fill opacity=0.8, draw opacity=1, text opacity=1, draw=white!80!black] at (rel axis cs: 0.85,0.15) {\shortstack[l]{
\ref{p1} Group 1 
}};

\node [fill opacity=0.8, draw opacity=1, text opacity=1, draw=white!80!black] at (rel axis cs: 0.85,0.85) {\shortstack[l]{
\ref{p2} Group 2 
}};

\end{tikzpicture}
    \caption{\textbf{ML model FL}: prediction error $\left( \hat{y}-\mathbb{E}[y] \right)$}
    \label{fig:ML_FL_prediction_error}
\end{subfigure}%
\vspace{10pt}
\begin{subfigure}[t]{0.44\textwidth}
    \centering
    \input{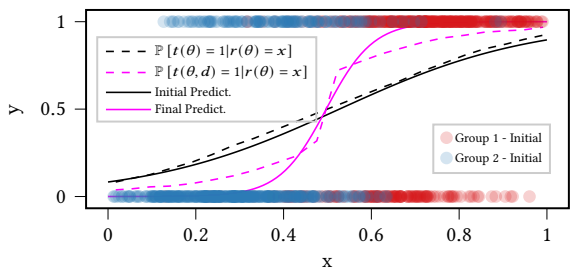}
    \caption{\textbf{Outcome FL}: initial distribution of $(X,Y)$, initial/final predictors, and initial/final outcome realization $t$}
    \label{fig:outcome_FL_predictions}
\end{subfigure}%
\hfill{}
\begin{subfigure}[t]{0.44\textwidth}
    \centering
    \input{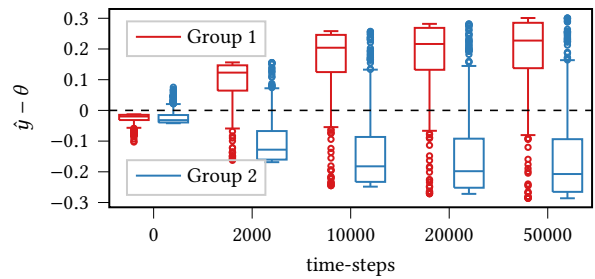}
    \caption{\textbf{Outcome FL}: prediction error with respect to the true, unobserved individual characteristics $\left( \hat{y}-\theta \right)$}
    \label{fig:outcome_FL_prediction_error}
\end{subfigure}%
\caption{Dynamic effects of different types of feedback loops (FL) acting on an RS pipeline for an online platform. 
Circles in the box plots denote outliers.
}
\label{fig:case_study_RS_results}
\end{figure}

\paragraph{Sampling Feedback Loop}

First, we look at a special case in which $d=0$ corresponds to not receiving any recommendation, leading to users leaving the platform.
Instead, when receiving $d=1$, users stay on the platform.
Initially, about 50\% of the active users on the platform are from group 1 and 50\% from group 2: $n_{G1}=496$ and $n_{G2}=504$.
Every time someone leaves the platform, a new user replaces them. To mimic users' homophily, the new user is drawn from group $1$ with probability $p=\frac{n_{G1}}{n}$ (else, from group $2$), i.e., the higher the percentage of users from group $1$ in the platform, the higher the probability the new user belongs to group $1$.
As can be seen in Fig.~\ref{fig:sampling_FL_users}, this phenomenon leads to the reduction of $n_{G2}$ from $504$ to $89$ individuals after 10,000 time-steps.
This distribution persists in future time-steps, suggesting that it is a (locally) stable equilibrium point of the dynamical system.
Group 2 is underrepresented on the platform in the long term with just 8.9\% of the platform users.
This corresponds to nuance (ii) of the representation bias as described in Section~\ref{sec:implications-algo-fairness}.
However, at the same time, nuance (iii) of the representation bias is present for both groups: since only those given $d=1$ stay on the platform, the sample of active users becomes less representative over time, i.e., only interested users (those with high values for $\theta$) stay on the platform (see Fig.~\ref{fig:sampling_FL_thetas}).
Notice that it is difficult to classify the sampling feedback loop as positive or negative in this case, as there is no initial representation bias against Group 2 that gets amplified by the loop.
The resulting biased equilibrium point is simply a property of the closed-loop dynamics.

\paragraph{Individual Feedback Loop}

An example of an individual feedback loop is when the recommended content influences the user's opinion $\theta$, which we model by letting the new opinion be a convex combination of the previous one and the recommended content.
Fig.~\ref{fig:individual_FL_thetas} shows that this results in a polarization of interests on the platform.
Namely, users with high initial interest (i.e., $\theta>0.5$) are more likely to be recommended the item and, as a result of this, their $\theta$ further increases over time, and vice versa for users with low initial interest.\footnote{
Notice that the underlying assumption is that any decision reinforces a user's opinion. This also means that users lose interest (i.e., $\theta$ decreases) if the relevant item ($d=0$) is not recommended.
}
Due to the initial difference in the distributions of $\theta$ (see Table~\ref{tab:case_study_initial_conditions}), historical bias increases.
Namely, results show bigger group-level disparities with a very high $\theta$ for group 1 users and a very low $\theta$ for group 2 users, on average.
The steady-state value reached by the trajectories in Fig.~\ref{fig:individual_FL_thetas} represents a biased stable equilibrium point of the closed-loop ML system in which the opinions are polarized.

\paragraph{Feature Feedback Loop}

Fig.~\ref{fig:feature_FL_thetas} shows the result of an example in which the content recommendation feeds back into the feature extraction block $x$ (rather than acting on the actual opinion $\theta$, as was the case in the previous example), thus forming a feature feedback loop.
Compared to all the other examples, there is no difference in the mean of the initial $\theta$ distribution across groups G1 and G2 $\left(\mu_{\theta,G1}=\mu_{\theta,G2}=0.5 \right)$.
However, initially, $x$ is a noisy observation of $\theta$ for both groups (i.e., there is measurement error $\sigma_{r,G1}=\sigma_{r,G2}=0.1$) and there is measurement bias on the feature observation of group 2 $\left( \mu_{r,G1}=0, \mu_{r,G2}=-0.2 \right)$.
This means that the interest of individuals of group 2 is systematically underestimated. 
Over time, the true interest of users is learned, since the feature feedback loop updates $x$ using new accurate information on the users' interests.
This reduces the measurement error, which can be seen by the reduced variance in Fig.~\ref{fig:feature_FL_thetas}.
Fig.~\ref{fig:feature_FL_thetas} further shows that the feature feedback loop reduces the measurement bias (which is measured as $x-\theta$) of group 2 over time, i.e., $x-\theta\sim0$ after $50,000$ time-steps.
Hence, the system converges to an unbiased stable equilibrium point, and the feature feedback loop leads to eliminating the measurement bias in this case.

\paragraph{ML Model Feedback Loop}

Next, we consider the case in which the content recommendation feeds back into the ML model.
For the initial training, we specify a high noise ($\sigma_{t_{train}}=1$) such that the model $f$ does not contain any information to predict $y$ from $x$, resulting in $\hat{y}=0.5$ for all individuals (see solid black line in Fig.~\ref{fig:ML_FL_predictions}).
Over time, the model observes new data and becomes more accurate, i.e., the predictor approximates the true outcome realization $t(\theta)$.

In this example, the training dataset $(X,Y)$ is enriched with feature-label pairs $\left(x,y\right)$ only if the relevant item was recommended to the user (i.e., if $d=1$, namely, partial feedback), assuming that the platform cannot measure the user's interest in an item that was not recommended.
This forms an ML model feedback loop.
The result is that after an initial period of exploration, the ML model quickly learns how to predict $y$ for individuals with large values of $x$, as those become the ones more likely to receive positive decisions.
Here, we measure the prediction error as $\hat{y}-\mathbb{E}[y]$, however, in the absence of any measurement error in the outcome realization ($t(\theta) = y$), $\theta$ is approximately equivalent to $\mathbb{E}[y]$ except for some noise, which is negligible for the average over a group of individuals.
As can be seen in Fig.~\ref{fig:ML_FL_prediction_error}, the prediction error quickly approaches 0 for G1, but the LR algorithm continues to perform poorly for G2 in the short to medium term.
In the long term, thanks to the noise in the observation of $x$\footnote{Namely, in a few cases, individuals from G2 have $x>0.5$ and therefore receive $d=1$. Thereby, the RS slowly explores the true distribution of group 2.} it eventually approaches 0 also for G2.

Retraining the ML model over time reduces the representation bias, nuance (ii) of the representation bias as described in Section~\ref{sec:implications-algo-fairness}.
However, it is due to the ML model feedback loop that the sample $(X,Y)$ becomes more representative of group 1 after just very few time-steps while taking much longer to reduce representation bias for group 2.

\paragraph{Outcome Feedback Loop}

Finally, we consider an example using the same initial conditions as in the sampling and individual feedback loops, but this time the RS's decision affects the outcome realization $t$.
Namely, the probability of the realized outcome to be $y=1$ increases/decreases by 20\% for positive/negative decisions, respectively.
This means that the realized outcomes $t(\theta,d)$ are more extreme than they would be if there were no outcome feedback loop (see dashed lines in Fig.~\ref{fig:outcome_FL_predictions}).
Despite starting with an unbiased ML model, over time, the retrained ML model approximates $t(\theta,d)$, i.e., the initial predictor is a much flatter sigmoid function compared to the final predictor.
Namely, the outcome feedback loop introduces a measurement bias on the realized outcome $y$ for both groups G1 and G2.
Thus, as is visible in Fig.~\ref{fig:outcome_FL_prediction_error}, the prediction error 
$\hat{y}-\theta$
diverges from 0 (as $\hat{y}$ predicts the realized outcome $y$ and not $\theta$) until it reaches a stable equilibrium point after approximately 10,000 time-steps (at approximately 0.2 and -0.2 for G1 and G2).
From the perspective of platform users, an outcome feedback loop can result in a situation in which one keeps receiving recommendations due to having clicked on similar
content in the past, despite not being interested in it.

\section{Related Work and Discussion}
\label{sec:related_work}

Along with providing a common system-theoretical framework for the growing literature on fairness in sequential decision-making (already discussed in Sec.~\ref{sec:introduction}), 
our aim is to pave the way for a fruitful collaboration with different communities working on, e.g., distribution shifts, adversarial machine learning, control theory, and optimal transport.
In what follows, we elaborate more on these research directions and on future interdisciplinary approaches.

\paragraph{Distribution Shifts}

Many works have investigated ML under different types of distribution shifts over time.
The problem of \textit{concept drift} is broadly defined as a shift of the target distribution over time~\cite{Gama2014ASurvey,Webb2016Characterizing}.
This is a rather broad definition, which includes distribution shifts due to exogenous effects, e.g., a pandemic or a financial crisis.
However, such shifts can be arbitrary and do not assume that feedback loops are present in general.
Recently, endogenous distribution shifts, i.e., target distribution shifts caused by the deployed prediction model, have been investigated more thoroughly.
The concept of \textit{performative predictions} acknowledges the fact that ML-based decision-making systems can affect the outcome they try to predict~\cite{Perdomo2020PerformativePrediction}.
The notion of performative stability, which is defined as a predictor that is not only calibrated against historical data but also against future outcomes that are produced by acting based on the prediction, is a possible solution that achieves a stable point for retraining~\cite{Perdomo2020PerformativePrediction}.
This stable point means that a model remains exactly the same if it is retrained on future outcomes.
Performative prediction is an umbrella term for a situation where ML-based decisions cause a shift in the outcome distribution.
However, this distribution shift can occur through any type of feedback loop we introduced in Section~\ref{ssec:Feedback-loops}.%
\footnote{
For certain feedback loops (e.g., sampling feedback loops), the distribution shift is delayed, and for others (such as the outcome feedback loop), the feedback effect occurs immediately, i.e., at the same time-step.
Furthermore, a distribution shift caused by an adversarial feature or adversarial individual feedback loop is a special case of performative prediction, which has been referred to as strategic classification~\cite{hardt2016StrategicClassification,Milli2019}.
}
As we showed in Section~\ref{sec:example}, these feedback loops have different properties and implications.
For example, changing a platform user's opinion (through an individual feedback loop) is very different from changing the individual's realized outcome (through an outcome feedback loop).
In all cases, the recommendation changes the individual's consumption. In the former case, this is caused by shaped preferences. In contrast, the decision-relevant individual attributes remain unaltered in the latter case.
More research is needed to study the effects of the specific feedback loops we classify on the concept of performative power~\cite{hardt2022performative}, which only considers shifting outcome distributions in its more general understanding as in the performative prediction literature.

\paragraph{Adversarial Machine Learning}

Adversarial ML studies attacks on ML algorithms and how they can be defended~\cite{Huang2011AdversarialMachineLearning,Vorobeychik2018AdversarialMachineLearning}.
The idea is that adversarial attacks are executed by an attacker who intends to influence some part of the ML pipeline, whereas the developer of the ML algorithm thwarts the attacker's objective. In contrast, feedback loops do not occur due to malicious external manipulation but are a direct consequence of the dynamics in sequential decision-making systems.
Yet, the outcomes of certain adversarial attacks are closely related to the feedback loops we classify in this paper. For example, data poisoning attacks are associated with ML model feedback loops in that they modify the data used for training.
Applying measures designed to counter adversarial attacks to deal with feedback loops in sequential decision-making systems represents an interesting avenue for future research -- for example, robust learning through data sub-sampling~\cite{Kearns1993MaliciousErrors} or trimmed optimization~\cite{Liu2017RobustLinReg} to counter ML model feedback loops. First results have shown that this becomes more complicated if the fairness of the decision-making systems is a concern~\cite{Xu2021RobustOrFair}.

\paragraph{Control Theory and Optimal Transport}

Control Theory provides the tools to drive dynamical systems towards a state with a desired behavior. These goals are reached by designing a \emph{controller} with the required corrective behavior.
Our framework provides a basis to interpret the decision-maker as a controller that can be purposely designed to mutually achieve desired performances and bias mitigation,
enabling the use of tools from Control Systems theory, such as Economic Model Predictive Control~\cite{EMPC} or Optimal Control~\cite{lewis2012optimal}.
Using these tools, one could incorporate fairness guarantees as constraints and performance guarantees, e.g., engagement on an online platform, as the objective function, possibly in a Regret minimization fashion~\cite{blum2007learning}. For the case of \textit{adversarial feedback loops}, one can think about the decision-maker and the external environment as the two players of a zero-sum game where Robust Control~\cite{zhou1998essentials} techniques find their natural expression. For example, referring to the selection process example in Section~\ref{sssec:Sampling_Feedback_Loop}, one can model the adversarial actions taken by the candidates as a disturbance for the decision-maker in achieving the best candidates selection.
When looking at bias mitigation techniques at the group level, tools from Optimal Transport can also come in handy \cite{Chen_optimal_transport}. Optimal Transport allows quantifying the violation of fairness constraints as the distance between the group's current distribution and a desired one \cite{OT:2020}. This tool would allow the design of a controller (decision maker) that drives the initial distribution towards an ideal one, fulfilling fairness constraints.

\section{Conclusion}

The output of ML-based decision-making systems, i.e., the decision, often affects various parts of the system itself, creating a so-called feedback loop.
Yet, ML evaluation techniques usually omit potentially important temporal dynamics~\cite{Chouldechova2018,mitchell2021algorithmic,Liu2018Delayed_Long_version} and taking feedback loops into account is crucial to avoid unintended consequences~\cite{Zhang2019GroupFairness,DAmour2020FairnessStudies,Liu2018Delayed_Long_version,sunbackfire}.
In this work, we build on dynamical systems theory to provide a general framework that sheds light on the different types of feedback loops that can occur throughout the ML pipeline.
We identify five distinct types of feedback loops, some of which can be classified as ``adversarial'' whenever the decision feeds back into the system as a consequence of some strategic action of the affected individual(s).
Furthermore, we associate the different types of feedback loops with the corresponding biases they affect, and we demonstrate these dynamics using a recommender system example.

By rigorously analyzing the ML pipeline, we believe that our framework is a necessary preliminary step towards (i) understanding the exact role of the feedback loops 
and (ii) shifting the research focus from short-sighted solutions that aim to identify and correct existing biases to a more forward-looking approach that seeks to anticipate and prevent biases in the long term.
First, providing a rigorous classification of feedback loops will pave the way for a systematic review of existing works in the ML literature and it will allow putting their results into the perspective of their assumptions (e.g., which types of feedback loops are considered and which are not).
Second, with the help of additional tools, e.g., dynamical systems and control theory, it will be possible to fully exploit the potential of our framework in the purposeful design of feedback loops, and for the development of effective long-term unfairness mitigation techniques.

\begin{acks}
We want to thank Kenny Joseph, Florian Dörfler, Sarah Dean, and the members of the Social Computing Group at the University of Zurich (Corinna Hertweck, Stefania Ionescu, Aleksandra Urman, Leonore Röseler, Azza Bouleimen, and Desheng Hu) for their feedback on an earlier version of this manuscript.
This work was supported by the University of Zurich, ETH Zurich, and NCCR Automation, a National Centre of Competence in Research, funded by the Swiss National Science Foundation (grant number $180545$).
\end{acks}

\bibliographystyle{ACM-Reference-Format}
\bibliography{NCCR_mendeley_references,NCCR_manual_references}


\begin{thebibliography}{77}


\ifx \showCODEN    \undefined \def \showCODEN     #1{\unskip}     \fi
\ifx \showDOI      \undefined \def \showDOI       #1{#1}\fi
\ifx \showISBNx    \undefined \def \showISBNx     #1{\unskip}     \fi
\ifx \showISBNxiii \undefined \def \showISBNxiii  #1{\unskip}     \fi
\ifx \showISSN     \undefined \def \showISSN      #1{\unskip}     \fi
\ifx \showLCCN     \undefined \def \showLCCN      #1{\unskip}     \fi
\ifx \shownote     \undefined \def \shownote      #1{#1}          \fi
\ifx \showarticletitle \undefined \def \showarticletitle #1{#1}   \fi
\ifx \showURL      \undefined \def \showURL       {\relax}        \fi
\providecommand\bibfield[2]{#2}
\providecommand\bibinfo[2]{#2}
\providecommand\natexlab[1]{#1}
\providecommand\showeprint[2][]{arXiv:#2}

\bibitem[Angwin et~al\mbox{.}(2016)]%
        {angwin2016machine}
\bibfield{author}{\bibinfo{person}{Julia Angwin}, \bibinfo{person}{Jeff
  Larson}, \bibinfo{person}{Surya Mattu}, {and} \bibinfo{person}{Lauren
  Kirchner}.} \bibinfo{year}{2016}\natexlab{}.
\newblock \showarticletitle{{Machine bias}}.
\newblock \bibinfo{journal}{\emph{ProPublica, May}} \bibinfo{volume}{23},
  \bibinfo{number}{2016} (\bibinfo{year}{2016}), \bibinfo{pages}{139--159}.
\newblock
\urldef\tempurl%
\url{https://www.propublica.org/article/machine-bias-risk-assessments-in-criminal-sentencing}
\showURL{%
\tempurl}


\bibitem[{\AA}str{\"{o}}m and Murray(2021)]%
        {aastrom2021feedback}
\bibfield{author}{\bibinfo{person}{Karl~Johan {\AA}str{\"{o}}m} {and}
  \bibinfo{person}{Richard~M Murray}.} \bibinfo{year}{2021}\natexlab{}.
\newblock \bibinfo{booktitle}{\emph{{Feedback systems: an introduction for
  scientists and engineers}}}.
\newblock \bibinfo{publisher}{Princeton university press}.
\newblock


\bibitem[Barocas et~al\mbox{.}(2019)]%
        {barocas-hardt-narayanan}
\bibfield{author}{\bibinfo{person}{Solon Barocas}, \bibinfo{person}{Moritz
  Hardt}, {and} \bibinfo{person}{Arvind Narayanan}.}
  \bibinfo{year}{2019}\natexlab{}.
\newblock \bibinfo{booktitle}{\emph{{Fairness and Machine Learning}}}.
\newblock \bibinfo{publisher}{fairmlbook.org}.
\newblock
\urldef\tempurl%
\url{http://www.fairmlbook.org}
\showURL{%
\tempurl}


\bibitem[Baumann et~al\mbox{.}(2022)]%
        {baumann2022sufficiency}
\bibfield{author}{\bibinfo{person}{Joachim Baumann}, \bibinfo{person}{Anikó
  Hann{\'{a}}k}, {and} \bibinfo{person}{Christoph Heitz}.}
  \bibinfo{year}{2022}\natexlab{}.
\newblock \showarticletitle{{Enforcing Group Fairness in Algorithmic Decision
  Making: Utility Maximization Under Sufficiency}}. In
  \bibinfo{booktitle}{\emph{Proceedings of the 2022 ACM Conference on Fairness,
  Accountability, and Transparency}} \emph{(\bibinfo{series}{FAccT '22})}.
  \bibinfo{publisher}{Association for Computing Machinery},
  \bibinfo{address}{New York, NY, USA}, \bibinfo{pages}{2315--2326}.
\newblock
\urldef\tempurl%
\url{https://doi.org/10.1145/3531146.3534645}
\showDOI{\tempurl}


\bibitem[Bechavod et~al\mbox{.}(2019)]%
        {Bechavod2019EqualFeedback}
\bibfield{author}{\bibinfo{person}{Yahav Bechavod}, \bibinfo{person}{Katrina
  Ligett}, \bibinfo{person}{Aaron Roth}, \bibinfo{person}{Bo Waggoner}, {and}
  \bibinfo{person}{Zhiwei~Steven Wu}.} \bibinfo{year}{2019}\natexlab{}.
\newblock \showarticletitle{{Equal opportunity in online classification with
  partial feedback}}.
\newblock \bibinfo{journal}{\emph{Advances in Neural Information Processing
  Systems}} \bibinfo{volume}{32}, \bibinfo{number}{NeurIPS}
  (\bibinfo{year}{2019}).
\newblock
\showISSN{10495258}


\bibitem[Berk et~al\mbox{.}(2021)]%
        {berk2021criminal}
\bibfield{author}{\bibinfo{person}{Richard Berk}, \bibinfo{person}{Hoda
  Heidari}, \bibinfo{person}{Shahin Jabbari}, \bibinfo{person}{Michael Kearns},
  {and} \bibinfo{person}{Aaron Roth}.} \bibinfo{year}{2021}\natexlab{}.
\newblock \showarticletitle{{Fairness in Criminal Justice Risk Assessments: The
  State of the Art}}.
\newblock \bibinfo{journal}{\emph{Sociological Methods {\&} Research}}
  \bibinfo{volume}{50}, \bibinfo{number}{1} (\bibinfo{year}{2021}),
  \bibinfo{pages}{3--44}.
\newblock
\urldef\tempurl%
\url{https://doi.org/10.1177/0049124118782533}
\showDOI{\tempurl}


\bibitem[Blum and Monsour(2007)]%
        {blum2007learning}
\bibfield{author}{\bibinfo{person}{Avrim Blum} {and} \bibinfo{person}{Yishay
  Monsour}.} \bibinfo{year}{2007}\natexlab{}.
\newblock \showarticletitle{Learning, regret minimization, and equilibria}.
\newblock  (\bibinfo{year}{2007}).
\newblock


\bibitem[Buolamwini and Gebru(2018)]%
        {pmlr-v81-buolamwini18a}
\bibfield{author}{\bibinfo{person}{Joy Buolamwini} {and}
  \bibinfo{person}{Timnit Gebru}.} \bibinfo{year}{2018}\natexlab{}.
\newblock \showarticletitle{{Gender Shades: Intersectional Accuracy Disparities
  in Commercial Gender Classification}}. In
  \bibinfo{booktitle}{\emph{Proceedings of the 1st Conference on Fairness,
  Accountability and Transparency}} \emph{(\bibinfo{series}{Proceedings of
  Machine Learning Research}, Vol.~\bibinfo{volume}{81})},
  \bibfield{editor}{\bibinfo{person}{Sorelle~A Friedler} {and}
  \bibinfo{person}{Christo Wilson}} (Eds.). \bibinfo{publisher}{PMLR},
  \bibinfo{pages}{77--91}.
\newblock
\urldef\tempurl%
\url{https://proceedings.mlr.press/v81/buolamwini18a.html}
\showURL{%
\tempurl}


\bibitem[Caton and Haas(2020)]%
        {caton2020fairness}
\bibfield{author}{\bibinfo{person}{Simon Caton} {and}
  \bibinfo{person}{Christian Haas}.} \bibinfo{year}{2020}\natexlab{}.
\newblock \showarticletitle{{Fairness in Machine Learning: A Survey}}.
\newblock  (\bibinfo{year}{2020}).
\newblock
\urldef\tempurl%
\url{http://arxiv.org/abs/2010.04053}
\showURL{%
\tempurl}


\bibitem[Chaney et~al\mbox{.}(2018)]%
        {chaney2018algorithmic}
\bibfield{author}{\bibinfo{person}{Allison J.~B. Chaney},
  \bibinfo{person}{Brandon~M. Stewart}, {and} \bibinfo{person}{Barbara~E.
  Engelhardt}.} \bibinfo{year}{2018}\natexlab{}.
\newblock \showarticletitle{{How algorithmic confounding in recommendation
  systems increases homogeneity and decreases utility}}.
\newblock  (\bibinfo{year}{2018}), \bibinfo{pages}{224--232}.
\newblock
\showISBNx{9781450359016}
\urldef\tempurl%
\url{https://doi.org/10.1145/3240323.3240370}
\showDOI{\tempurl}


\bibitem[Chen et~al\mbox{.}(2021)]%
        {Chen_optimal_transport}
\bibfield{author}{\bibinfo{person}{Yongxin Chen}, \bibinfo{person}{Tryphon~T.
  Georgiou}, {and} \bibinfo{person}{Michele Pavon}.}
  \bibinfo{year}{2021}\natexlab{}.
\newblock \showarticletitle{Optimal Transport in Systems and Control}.
\newblock \bibinfo{journal}{\emph{Annual Review of Control, Robotics, and
  Autonomous Systems}} \bibinfo{volume}{4}, \bibinfo{number}{1}
  (\bibinfo{year}{2021}), \bibinfo{pages}{89--113}.
\newblock
\urldef\tempurl%
\url{https://doi.org/10.1146/annurev-control-070220-100858}
\showDOI{\tempurl}


\bibitem[Chiappa et~al\mbox{.}(2020)]%
        {OT:2020}
\bibfield{author}{\bibinfo{person}{Silvia Chiappa}, \bibinfo{person}{Ray
  Jiang}, \bibinfo{person}{Tom Stepleton}, \bibinfo{person}{Aldo Pacchiano},
  \bibinfo{person}{Heinrich Jiang}, {and} \bibinfo{person}{John Aslanides}.}
  \bibinfo{year}{2020}\natexlab{}.
\newblock \showarticletitle{{A General Approach to Fairness with Optimal
  Transport}}.
\newblock \bibinfo{journal}{\emph{The 34th AAAI Conference on Artificial
  Intelligence}} (\bibinfo{year}{2020}).
\newblock
\urldef\tempurl%
\url{https://ojs.aaai.org/index.php/AAAI/article/view/5771}
\showURL{%
\tempurl}


\bibitem[Chouldechova and Roth(2018)]%
        {Chouldechova2018}
\bibfield{author}{\bibinfo{person}{Alexandra Chouldechova} {and}
  \bibinfo{person}{Aaron Roth}.} \bibinfo{year}{2018}\natexlab{}.
\newblock \showarticletitle{{The Frontiers of Fairness in Machine Learning}}.
\newblock  (\bibinfo{year}{2018}), \bibinfo{pages}{1--13}.
\newblock
\urldef\tempurl%
\url{http://arxiv.org/abs/1810.08810}
\showURL{%
\tempurl}


\bibitem[Chouldechova and Roth(2020)]%
        {Chouldechova2020}
\bibfield{author}{\bibinfo{person}{Alexandra Chouldechova} {and}
  \bibinfo{person}{Aaron Roth}.} \bibinfo{year}{2020}\natexlab{}.
\newblock \showarticletitle{{A Snapshot of the Frontiers of Fairness in Machine
  Learning}}.
\newblock \bibinfo{journal}{\emph{Commun. ACM}} \bibinfo{volume}{63},
  \bibinfo{number}{5} (\bibinfo{date}{4} \bibinfo{year}{2020}),
  \bibinfo{pages}{82–89}.
\newblock
\showISSN{0001-0782}
\urldef\tempurl%
\url{https://doi.org/10.1145/3376898}
\showDOI{\tempurl}


\bibitem[Corbett-Davies et~al\mbox{.}(2017)]%
        {10.1145/3097983.3098095}
\bibfield{author}{\bibinfo{person}{Sam Corbett-Davies}, \bibinfo{person}{Emma
  Pierson}, \bibinfo{person}{Avi Feller}, \bibinfo{person}{Sharad Goel}, {and}
  \bibinfo{person}{Aziz Huq}.} \bibinfo{year}{2017}\natexlab{}.
\newblock \showarticletitle{{Algorithmic Decision Making and the Cost of
  Fairness}}. In \bibinfo{booktitle}{\emph{Proceedings of the 23rd ACM SIGKDD
  International Conference on Knowledge Discovery and Data Mining}}
  \emph{(\bibinfo{series}{KDD '17})}. \bibinfo{publisher}{Association for
  Computing Machinery}, \bibinfo{address}{New York, NY, USA},
  \bibinfo{pages}{797–806}.
\newblock
\showISBNx{9781450348874}
\urldef\tempurl%
\url{https://doi.org/10.1145/3097983.3098095}
\showDOI{\tempurl}


\bibitem[Crawford(2016)]%
        {crawford2016artificial}
\bibfield{author}{\bibinfo{person}{Kate Crawford}.}
  \bibinfo{year}{2016}\natexlab{}.
\newblock \showarticletitle{{Artificial intelligence’s white guy problem}}.
\newblock \bibinfo{journal}{\emph{The New York Times}} \bibinfo{volume}{25},
  \bibinfo{number}{06} (\bibinfo{year}{2016}).
\newblock


\bibitem[D'Amour et~al\mbox{.}(2020)]%
        {DAmour2020FairnessStudies}
\bibfield{author}{\bibinfo{person}{Alexander D'Amour}, \bibinfo{person}{Hansa
  Srinivasan}, \bibinfo{person}{James Atwood}, \bibinfo{person}{Pallavi
  Baljekar}, \bibinfo{person}{D. Sculley}, {and} \bibinfo{person}{Yoni
  Halpern}.} \bibinfo{year}{2020}\natexlab{}.
\newblock \showarticletitle{{Fairness is not static: Deeper understanding of
  long term fairness via simulation studies}}.
\newblock \bibinfo{journal}{\emph{FAT* 2020 - Proceedings of the 2020
  Conference on Fairness, Accountability, and Transparency}}
  (\bibinfo{year}{2020}), \bibinfo{pages}{525--534}.
\newblock
\showISBNx{9781450369367}
\urldef\tempurl%
\url{https://doi.org/10.1145/3351095.3372878}
\showDOI{\tempurl}


\bibitem[Dawes et~al\mbox{.}(1989)]%
        {Dawes1989}
\bibfield{author}{\bibinfo{person}{Robyn~M Dawes}, \bibinfo{person}{David
  Faust}, {and} \bibinfo{person}{Paul~E Meehl}.}
  \bibinfo{year}{1989}\natexlab{}.
\newblock \showarticletitle{{Clinical Versus Actuarial Judgment}}.
\newblock \bibinfo{journal}{\emph{Science}} \bibinfo{volume}{243},
  \bibinfo{number}{4899} (\bibinfo{year}{1989}), \bibinfo{pages}{1668--1674}.
\newblock
\urldef\tempurl%
\url{https://doi.org/10.1126/science.2648573}
\showDOI{\tempurl}


\bibitem[Dobbe et~al\mbox{.}(2018)]%
        {dobbe2018broader}
\bibfield{author}{\bibinfo{person}{Roel Dobbe}, \bibinfo{person}{Sarah Dean},
  \bibinfo{person}{Thomas Gilbert}, {and} \bibinfo{person}{Nitin Kohli}.}
  \bibinfo{year}{2018}\natexlab{}.
\newblock \showarticletitle{A broader view on bias in automated
  decision-making: Reflecting on epistemology and dynamics}.
\newblock \bibinfo{journal}{\emph{arXiv preprint arXiv:1807.00553}}
  (\bibinfo{year}{2018}).
\newblock


\bibitem[Ellis et~al\mbox{.}(2014)]%
        {EMPC}
\bibfield{author}{\bibinfo{person}{Matthew Ellis}, \bibinfo{person}{Helen
  Durand}, {and} \bibinfo{person}{Panagiotis~D. Christofides}.}
  \bibinfo{year}{2014}\natexlab{}.
\newblock \showarticletitle{{A tutorial review of economic model predictive
  control methods}}.
\newblock \bibinfo{journal}{\emph{Journal of Process Control}}
  \bibinfo{volume}{24}, \bibinfo{number}{8} (\bibinfo{year}{2014}).
\newblock
\showISSN{0959-1524}
\urldef\tempurl%
\url{http://dx.doi.org/10.1016/j.jprocont.2014.03.01}
\showURL{%
\tempurl}


\bibitem[Elzayn et~al\mbox{.}(2019)]%
        {Elzayn2019FairProblems}
\bibfield{author}{\bibinfo{person}{Hadi Elzayn}, \bibinfo{person}{Michael
  Kearns}, \bibinfo{person}{Shahin Jabbari}, \bibinfo{person}{Seth Neel},
  \bibinfo{person}{Zachary Schutzman}, \bibinfo{person}{Christopher Jung},
  {and} \bibinfo{person}{Aaron Roth}.} \bibinfo{year}{2019}\natexlab{}.
\newblock \showarticletitle{{Fair algorithms for learning in allocation
  problems}}.
\newblock \bibinfo{journal}{\emph{FAT* 2019 - Proceedings of the 2019
  Conference on Fairness, Accountability, and Transparency}}
  (\bibinfo{year}{2019}), \bibinfo{pages}{170--179}.
\newblock
\showISBNx{9781450361255}
\urldef\tempurl%
\url{https://doi.org/10.1145/3287560.3287571}
\showDOI{\tempurl}


\bibitem[Ensign et~al\mbox{.}(2018b)]%
        {Ensign2018DecisionPrediction}
\bibfield{author}{\bibinfo{person}{Danielle Ensign}, \bibinfo{person}{Sorelle~A
  Friedler}, \bibinfo{person}{Scott Neville}, \bibinfo{person}{Carlos
  Scheidegger}, \bibinfo{person}{Suresh Venkatasubramanian},
  \bibinfo{person}{Mehryar Mohri}, {and} \bibinfo{person}{Karthik Sridharan}.}
  \bibinfo{year}{2018}\natexlab{b}.
\newblock \showarticletitle{{Decision making with limited feedback: Error
  bounds for predictive policing and recidivism prediction}}.
\newblock \bibinfo{journal}{\emph{Proceedings of Machine Learning Research}}
  \bibinfo{volume}{83} (\bibinfo{year}{2018}), \bibinfo{pages}{1--9}.
\newblock


\bibitem[Ensign et~al\mbox{.}(2018a)]%
        {Ensign2018RunawayPolicing}
\bibfield{author}{\bibinfo{person}{Danielle Ensign}, \bibinfo{person}{Sorelle~A
  Friedler}, \bibinfo{person}{Scott Neville}, \bibinfo{person}{Carlos
  Scheidegger}, \bibinfo{person}{Suresh Venkatasubramanian}, {and}
  \bibinfo{person}{Christo Wilson}.} \bibinfo{year}{2018}\natexlab{a}.
\newblock \showarticletitle{{Runaway Feedback Loops in Predictive Policing}}.
  In \bibinfo{booktitle}{\emph{Proceedings of Machine Learning Research}},
  Vol.~\bibinfo{volume}{81}. \bibinfo{pages}{1--12}.
\newblock
\urldef\tempurl%
\url{https://github.com/algofairness/}
\showURL{%
\tempurl}


\bibitem[Friedler et~al\mbox{.}(2016)]%
        {Friedler2016}
\bibfield{author}{\bibinfo{person}{Sorelle~A Friedler}, \bibinfo{person}{Carlos
  Scheidegger}, {and} \bibinfo{person}{Suresh Venkatasubramanian}.}
  \bibinfo{year}{2016}\natexlab{}.
\newblock \bibinfo{title}{{On the (im)possibility of fairness}}.
\newblock
\newblock
\urldef\tempurl%
\url{https://arxiv.org/abs/1609.07236}
\showURL{%
\tempurl}


\bibitem[Fuster et~al\mbox{.}(2022)]%
        {Fuster2022PredictablyMarkets}
\bibfield{author}{\bibinfo{person}{Andreas Fuster}, \bibinfo{person}{Paul
  Goldsmith-Pinkham}, \bibinfo{person}{Tarun Ramadorai}, {and}
  \bibinfo{person}{Ansgar Walther}.} \bibinfo{year}{2022}\natexlab{}.
\newblock \showarticletitle{{Predictably Unequal? The Effects of Machine
  Learning on Credit Markets}}.
\newblock \bibinfo{journal}{\emph{Journal of Finance}} \bibinfo{volume}{77},
  \bibinfo{number}{1} (\bibinfo{year}{2022}), \bibinfo{pages}{5--47}.
\newblock
\showISSN{15406261}
\urldef\tempurl%
\url{https://doi.org/10.1111/jofi.13090}
\showDOI{\tempurl}


\bibitem[Gama et~al\mbox{.}(2014)]%
        {Gama2014ASurvey}
\bibfield{author}{\bibinfo{person}{João Gama}, \bibinfo{person}{Indrundefined
  {\v{Z}}liobaitundefined}, \bibinfo{person}{Albert Bifet},
  \bibinfo{person}{Mykola Pechenizkiy}, {and} \bibinfo{person}{Abdelhamid
  Bouchachia}.} \bibinfo{year}{2014}\natexlab{}.
\newblock \showarticletitle{{A Survey on Concept Drift Adaptation}}.
\newblock \bibinfo{journal}{\emph{ACM Comput. Surv.}} \bibinfo{volume}{46},
  \bibinfo{number}{4} (\bibinfo{date}{3} \bibinfo{year}{2014}).
\newblock
\showISSN{0360-0300}
\urldef\tempurl%
\url{https://doi.org/10.1145/2523813}
\showDOI{\tempurl}


\bibitem[Grove et~al\mbox{.}(2000)]%
        {Grove2000}
\bibfield{author}{\bibinfo{person}{W~M Grove}, \bibinfo{person}{D~H Zald},
  \bibinfo{person}{B~S Lebow}, \bibinfo{person}{B~E Snitz}, {and}
  \bibinfo{person}{C Nelson}.} \bibinfo{year}{2000}\natexlab{}.
\newblock \showarticletitle{{Clinical versus mechanical prediction: a
  meta-analysis.}}
\newblock \bibinfo{journal}{\emph{Psychological assessment}}
  \bibinfo{volume}{12}, \bibinfo{number}{1} (\bibinfo{date}{3}
  \bibinfo{year}{2000}), \bibinfo{pages}{19--30}.
\newblock
\showISSN{1040-3590 (Print)}


\bibitem[Hardt et~al\mbox{.}(2022)]%
        {hardt2022performative}
\bibfield{author}{\bibinfo{person}{Moritz Hardt}, \bibinfo{person}{Meena
  Jagadeesan}, {and} \bibinfo{person}{Celestine Mendler-D{\"{u}}nner}.}
  \bibinfo{year}{2022}\natexlab{}.
\newblock \showarticletitle{{Performative Power}}. In
  \bibinfo{booktitle}{\emph{Advances in Neural Information Processing
  Systems}}, \bibfield{editor}{\bibinfo{person}{Alice~H Oh},
  \bibinfo{person}{Alekh Agarwal}, \bibinfo{person}{Danielle Belgrave}, {and}
  \bibinfo{person}{Kyunghyun Cho}} (Eds.).
\newblock
\urldef\tempurl%
\url{https://doi.org/10.48550/ARXIV.2203.17232}
\showDOI{\tempurl}


\bibitem[Hardt et~al\mbox{.}(2016a)]%
        {hardt2016StrategicClassification}
\bibfield{author}{\bibinfo{person}{Moritz Hardt}, \bibinfo{person}{Nimrod
  Megiddo}, \bibinfo{person}{Christos Papadimitriou}, {and}
  \bibinfo{person}{Mary Wootters}.} \bibinfo{year}{2016}\natexlab{a}.
\newblock \showarticletitle{{Strategic Classification}}. In
  \bibinfo{booktitle}{\emph{Proceedings of the 2016 ACM Conference on
  Innovations in Theoretical Computer Science}} \emph{(\bibinfo{series}{ITCS
  '16})}. \bibinfo{publisher}{Association for Computing Machinery},
  \bibinfo{address}{New York, NY, USA}, \bibinfo{pages}{111–122}.
\newblock
\showISBNx{9781450340571}
\urldef\tempurl%
\url{https://doi.org/10.1145/2840728.2840730}
\showDOI{\tempurl}


\bibitem[Hardt et~al\mbox{.}(2016b)]%
        {hardt2016equality}
\bibfield{author}{\bibinfo{person}{Moritz Hardt}, \bibinfo{person}{Eric Price},
  {and} \bibinfo{person}{Nathan Srebro}.} \bibinfo{year}{2016}\natexlab{b}.
\newblock \showarticletitle{{Equality of opportunity in supervised learning}}.
  In \bibinfo{booktitle}{\emph{Advances in Neural Information Processing
  Systems}} \emph{(\bibinfo{series}{NIPS'16})}. \bibinfo{publisher}{Curran
  Associates Inc.}, \bibinfo{address}{Red Hook, NY, USA},
  \bibinfo{pages}{3323--3331}.
\newblock
\showISBNx{9781510838819}
\showISSN{10495258}


\bibitem[Harwell(2018)]%
        {AmazonAlexa2022}
\bibfield{author}{\bibinfo{person}{Drew Harwell}.}
  \bibinfo{year}{2018}\natexlab{}.
\newblock \bibinfo{title}{{Amazon’s Alexa and Google Home show accent bias,
  with Chinese and Spanish hardest to understand}}.
\newblock
\newblock
\urldef\tempurl%
\url{https://www.scmp.com/magazines/post-magazine/long-reads/article/2156455/amazons-alexa-and-google-home-show-accent-bias}
\showURL{%
\tempurl}


\bibitem[Hashimoto et~al\mbox{.}(2018)]%
        {pmlr-v80-hashimoto18a}
\bibfield{author}{\bibinfo{person}{Tatsunori Hashimoto}, \bibinfo{person}{Megha
  Srivastava}, \bibinfo{person}{Hongseok Namkoong}, {and}
  \bibinfo{person}{Percy Liang}.} \bibinfo{year}{2018}\natexlab{}.
\newblock \showarticletitle{{Fairness Without Demographics in Repeated Loss
  Minimization}}. In \bibinfo{booktitle}{\emph{Proceedings of the 35th
  International Conference on Machine Learning}}
  \emph{(\bibinfo{series}{Proceedings of Machine Learning Research},
  Vol.~\bibinfo{volume}{80})}, \bibfield{editor}{\bibinfo{person}{Jennifer Dy}
  {and} \bibinfo{person}{Andreas Krause}} (Eds.). \bibinfo{publisher}{PMLR},
  \bibinfo{pages}{1929--1938}.
\newblock
\urldef\tempurl%
\url{https://proceedings.mlr.press/v80/hashimoto18a.html}
\showURL{%
\tempurl}


\bibitem[Heidari et~al\mbox{.}(2019)]%
        {Heidari2019OnLearning}
\bibfield{author}{\bibinfo{person}{Hoda Heidari}, \bibinfo{person}{Vedant
  Nanda}, {and} \bibinfo{person}{Krishna~P. Gummadi}.}
  \bibinfo{year}{2019}\natexlab{}.
\newblock \showarticletitle{{On the Long-term Impact of Algorithmic Decision
  Policies: Effort unfairness and feature segregation through social
  learning}}.
\newblock \bibinfo{journal}{\emph{36th International Conference on Machine
  Learning, ICML 2019}}  \bibinfo{volume}{2019-June} (\bibinfo{year}{2019}),
  \bibinfo{pages}{4787--4796}.
\newblock
\showISBNx{9781510886988}


\bibitem[Hertweck et~al\mbox{.}(2021)]%
        {Hertweck2021statistical-parity}
\bibfield{author}{\bibinfo{person}{Corinna Hertweck},
  \bibinfo{person}{Christoph Heitz}, {and} \bibinfo{person}{Michele Loi}.}
  \bibinfo{year}{2021}\natexlab{}.
\newblock \showarticletitle{{On the Moral Justification of Statistical
  Parity}}. In \bibinfo{booktitle}{\emph{Proceedings of the 2021 ACM Conference
  on Fairness, Accountability, and Transparency}} \emph{(\bibinfo{series}{FAccT
  '21})}. \bibinfo{publisher}{Association for Computing Machinery},
  \bibinfo{address}{New York, NY, USA}, \bibinfo{pages}{747–757}.
\newblock
\showISBNx{9781450383097}
\urldef\tempurl%
\url{https://doi.org/10.1145/3442188.3445936}
\showDOI{\tempurl}


\bibitem[Hu and Chen(2018)]%
        {Hu2018AMarket}
\bibfield{author}{\bibinfo{person}{Lily Hu} {and} \bibinfo{person}{Yiling
  Chen}.} \bibinfo{year}{2018}\natexlab{}.
\newblock \showarticletitle{{A short-term intervention for long-term fairness
  in the labor market}}.
\newblock \bibinfo{journal}{\emph{The Web Conference 2018 - Proceedings of the
  World Wide Web Conference, WWW 2018}}  \bibinfo{volume}{2}
  (\bibinfo{year}{2018}), \bibinfo{pages}{1389--1398}.
\newblock
\showISBNx{9781450356398}
\urldef\tempurl%
\url{https://doi.org/10.1145/3178876.3186044}
\showDOI{\tempurl}


\bibitem[Hu et~al\mbox{.}(2019)]%
        {Hu2018TheManipulation}
\bibfield{author}{\bibinfo{person}{Lily Hu}, \bibinfo{person}{Nicole
  Immorlica}, {and} \bibinfo{person}{Jennifer~Wortman Vaughan}.}
  \bibinfo{year}{2019}\natexlab{}.
\newblock \showarticletitle{{The Disparate Effects of Strategic Manipulation}}.
  In \bibinfo{booktitle}{\emph{Proceedings of the Conference on Fairness,
  Accountability, and Transparency}} \emph{(\bibinfo{series}{FAT* '19})}.
  \bibinfo{publisher}{Association for Computing Machinery},
  \bibinfo{address}{New York, NY, USA}, \bibinfo{pages}{259–268}.
\newblock
\showISBNx{9781450361255}
\urldef\tempurl%
\url{https://doi.org/10.1145/3287560.3287597}
\showDOI{\tempurl}


\bibitem[Huang et~al\mbox{.}(2011)]%
        {Huang2011AdversarialMachineLearning}
\bibfield{author}{\bibinfo{person}{Ling Huang}, \bibinfo{person}{Anthony~D
  Joseph}, \bibinfo{person}{Blaine Nelson}, \bibinfo{person}{Benjamin I~P
  Rubinstein}, {and} \bibinfo{person}{J~D Tygar}.}
  \bibinfo{year}{2011}\natexlab{}.
\newblock \showarticletitle{{Adversarial Machine Learning}}. In
  \bibinfo{booktitle}{\emph{Proceedings of the 4th ACM Workshop on Security and
  Artificial Intelligence}} \emph{(\bibinfo{series}{AISec '11})}.
  \bibinfo{publisher}{Association for Computing Machinery},
  \bibinfo{address}{New York, NY, USA}, \bibinfo{pages}{43–58}.
\newblock
\showISBNx{9781450310031}
\urldef\tempurl%
\url{https://doi.org/10.1145/2046684.2046692}
\showDOI{\tempurl}


\bibitem[John~D.(2000)]%
        {JDS:00}
\bibfield{author}{\bibinfo{person}{Sterman John~D.}}
  \bibinfo{year}{2000}\natexlab{}.
\newblock \bibinfo{booktitle}{\emph{{Business Dynamics: Systems Thinking and
  Modeling for a Complex World}}}.
\newblock \bibinfo{publisher}{McGraw-Hill Education}.
\newblock
\showISBNx{ISBN 978-0-07-238915-9}


\bibitem[Kearns and Li(1993)]%
        {Kearns1993MaliciousErrors}
\bibfield{author}{\bibinfo{person}{Michael Kearns} {and} \bibinfo{person}{Ming
  Li}.} \bibinfo{year}{1993}\natexlab{}.
\newblock \showarticletitle{{Learning in the Presence of Malicious Errors}}.
\newblock \bibinfo{journal}{\emph{SIAM J. Comput.}} \bibinfo{volume}{22},
  \bibinfo{number}{4} (\bibinfo{year}{1993}), \bibinfo{pages}{807--837}.
\newblock
\urldef\tempurl%
\url{https://doi.org/10.1137/0222052}
\showDOI{\tempurl}


\bibitem[Kearns and Roth(2019)]%
        {Kearns2019EthicalAlgorithm}
\bibfield{author}{\bibinfo{person}{Michael Kearns} {and} \bibinfo{person}{Aaron
  Roth}.} \bibinfo{year}{2019}\natexlab{}.
\newblock \bibinfo{booktitle}{\emph{{The Ethical Algorithm: The Science of
  Socially Aware Algorithm Design}}}.
\newblock \bibinfo{publisher}{Oxford University Press, Inc.},
  \bibinfo{address}{USA}.
\newblock
\showISBNx{0190948205}


\bibitem[Kleinberg et~al\mbox{.}(2017)]%
        {Kleinberg2017HumanDecisionsMachinePredictions}
\bibfield{author}{\bibinfo{person}{Jon Kleinberg}, \bibinfo{person}{Himabindu
  Lakkaraju}, \bibinfo{person}{Jure Leskovec}, \bibinfo{person}{Jens Ludwig},
  {and} \bibinfo{person}{Sendhil Mullainathan}.}
  \bibinfo{year}{2017}\natexlab{}.
\newblock \bibinfo{booktitle}{\emph{{Human Decisions and Machine
  Predictions}}}.
\newblock \bibinfo{type}{{T}echnical {R}eport} 23180.
  \bibinfo{institution}{National Bureau of Economic Research}.
\newblock
\urldef\tempurl%
\url{https://doi.org/10.3386/w23180}
\showDOI{\tempurl}


\bibitem[Kleinberg et~al\mbox{.}(2018)]%
        {Kleinberg2018}
\bibfield{author}{\bibinfo{person}{Jon Kleinberg}, \bibinfo{person}{Jens
  Ludwig}, \bibinfo{person}{Sendhil Mullainathan}, {and}
  \bibinfo{person}{Ashesh Rambachan}.} \bibinfo{year}{2018}\natexlab{}.
\newblock \showarticletitle{{Algorithmic Fairness}}.
\newblock \bibinfo{journal}{\emph{AEA Papers and Proceedings}}
  \bibinfo{volume}{108} (\bibinfo{date}{5} \bibinfo{year}{2018}),
  \bibinfo{pages}{22--27}.
\newblock
\urldef\tempurl%
\url{https://doi.org/10.1257/pandp.20181018}
\showDOI{\tempurl}


\bibitem[Kleinberg and Raghavan(2020)]%
        {Kleinberg2020HowStrategically}
\bibfield{author}{\bibinfo{person}{Jon Kleinberg} {and} \bibinfo{person}{Manish
  Raghavan}.} \bibinfo{year}{2020}\natexlab{}.
\newblock \showarticletitle{{How Do Classifiers Induce Agents to Invest Effort
  Strategically?}}
\newblock \bibinfo{journal}{\emph{ACM Transactions on Economics and
  Computation}} \bibinfo{volume}{8}, \bibinfo{number}{4}
  (\bibinfo{year}{2020}).
\newblock
\showISSN{21678383}
\urldef\tempurl%
\url{https://doi.org/10.1145/3417742}
\showDOI{\tempurl}


\bibitem[Lewis et~al\mbox{.}(2012)]%
        {lewis2012optimal}
\bibfield{author}{\bibinfo{person}{Frank~L Lewis}, \bibinfo{person}{Draguna
  Vrabie}, {and} \bibinfo{person}{Vassilis~L Syrmos}.}
  \bibinfo{year}{2012}\natexlab{}.
\newblock \bibinfo{booktitle}{\emph{Optimal control}}.
\newblock \bibinfo{publisher}{John Wiley \& Sons}.
\newblock


\bibitem[Liu et~al\mbox{.}(2017)]%
        {Liu2017RobustLinReg}
\bibfield{author}{\bibinfo{person}{Chang Liu}, \bibinfo{person}{Bo Li},
  \bibinfo{person}{Yevgeniy Vorobeychik}, {and} \bibinfo{person}{Alina Oprea}.}
  \bibinfo{year}{2017}\natexlab{}.
\newblock \showarticletitle{{Robust Linear Regression Against Training Data
  Poisoning}}. In \bibinfo{booktitle}{\emph{Proceedings of the 10th ACM
  Workshop on Artificial Intelligence and Security}}
  \emph{(\bibinfo{series}{AISec '17})}. \bibinfo{publisher}{Association for
  Computing Machinery}, \bibinfo{address}{New York, NY, USA},
  \bibinfo{pages}{91–102}.
\newblock
\showISBNx{9781450352024}
\urldef\tempurl%
\url{https://doi.org/10.1145/3128572.3140447}
\showDOI{\tempurl}


\bibitem[Liu et~al\mbox{.}(2018a)]%
        {Liu2018Delayed_Long_version}
\bibfield{author}{\bibinfo{person}{Lydia~T Liu}, \bibinfo{person}{Sarah Dean},
  \bibinfo{person}{Esther Rolf}, \bibinfo{person}{Max Simchowitz}, {and}
  \bibinfo{person}{Moritz Hardt}.} \bibinfo{year}{2018}\natexlab{a}.
\newblock \showarticletitle{{Delayed Impact of Fair Machine Learning}}. In
  \bibinfo{booktitle}{\emph{Proceedings of the 35th International Conference on
  Machine Learning}} \emph{(\bibinfo{series}{Proceedings of Machine Learning
  Research}, Vol.~\bibinfo{volume}{80})},
  \bibfield{editor}{\bibinfo{person}{Jennifer Dy} {and}
  \bibinfo{person}{Andreas Krause}} (Eds.). \bibinfo{publisher}{PMLR},
  \bibinfo{pages}{3150--3158}.
\newblock
\urldef\tempurl%
\url{https://proceedings.mlr.press/v80/liu18c.html}
\showURL{%
\tempurl}


\bibitem[Liu et~al\mbox{.}(2018b)]%
        {Liu2018}
\bibfield{author}{\bibinfo{person}{Lydia~T Liu}, \bibinfo{person}{Sarah Dean},
  \bibinfo{person}{Esther Rolf}, \bibinfo{person}{Max Simchowitz}, {and}
  \bibinfo{person}{Moritz Hardt}.} \bibinfo{year}{2018}\natexlab{b}.
\newblock \showarticletitle{{Delayed Impact of Fair Machine Learning}}. In
  \bibinfo{booktitle}{\emph{Proceedings of the 35th International Conference on
  Machine Learning}}.
\newblock
\urldef\tempurl%
\url{https://proceedings.mlr.press/v80/liu18c.html}
\showURL{%
\tempurl}


\bibitem[Liu et~al\mbox{.}(2020)]%
        {Liu2020thedisparate}
\bibfield{author}{\bibinfo{person}{Lydia~T. Liu}, \bibinfo{person}{Adam~Tauman
  Kalai}, \bibinfo{person}{Ashia Wilson}, \bibinfo{person}{Christian Borgs},
  \bibinfo{person}{Nika Haghtalab}, {and} \bibinfo{person}{Jennifer Chayes}.}
  \bibinfo{year}{2020}\natexlab{}.
\newblock \showarticletitle{{The disparate equilibria of algorithmic decision
  making when individuals invest rationally}}.
\newblock \bibinfo{journal}{\emph{FAT* 2020 - Proceedings of the 2020
  Conference on Fairness, Accountability, and Transparency}}
  (\bibinfo{year}{2020}), \bibinfo{pages}{381--391}.
\newblock
\showISBNx{9781450369367}
\urldef\tempurl%
\url{https://doi.org/10.1145/3351095.3372861}
\showDOI{\tempurl}


\bibitem[Lum and Isaac(2016)]%
        {Lum2016Topredict}
\bibfield{author}{\bibinfo{person}{Kristian Lum} {and} \bibinfo{person}{William
  Isaac}.} \bibinfo{year}{2016}\natexlab{}.
\newblock \showarticletitle{{To predict and serve?}}
\newblock \bibinfo{journal}{\emph{Significance}} \bibinfo{volume}{13},
  \bibinfo{number}{5} (\bibinfo{year}{2016}), \bibinfo{pages}{14--19}.
\newblock
\urldef\tempurl%
\url{https://doi.org/10.1111/j.1740-9713.2016.00960.x}
\showDOI{\tempurl}


\bibitem[Mehrabi et~al\mbox{.}(2021)]%
        {mehrabi2021survey}
\bibfield{author}{\bibinfo{person}{Ninareh Mehrabi}, \bibinfo{person}{Fred
  Morstatter}, \bibinfo{person}{Nripsuta Saxena}, \bibinfo{person}{Kristina
  Lerman}, {and} \bibinfo{person}{Aram Galstyan}.}
  \bibinfo{year}{2021}\natexlab{}.
\newblock \showarticletitle{{A Survey on Bias and Fairness in Machine
  Learning}}.
\newblock \bibinfo{journal}{\emph{ACM Comput. Surv.}} \bibinfo{volume}{54},
  \bibinfo{number}{6} (\bibinfo{date}{7} \bibinfo{year}{2021}).
\newblock
\showISSN{0360-0300}
\urldef\tempurl%
\url{https://doi.org/10.1145/3457607}
\showDOI{\tempurl}


\bibitem[Milli et~al\mbox{.}(2019)]%
        {Milli2019}
\bibfield{author}{\bibinfo{person}{Smitha Milli}, \bibinfo{person}{John
  Miller}, \bibinfo{person}{Anca~D. Dragan}, {and} \bibinfo{person}{Moritz
  Hardt}.} \bibinfo{year}{2019}\natexlab{}.
\newblock \showarticletitle{{The social cost of strategic classification}}.
\newblock \bibinfo{journal}{\emph{FAT* 2019 - Proceedings of the 2019
  Conference on Fairness, Accountability, and Transparency}}
  (\bibinfo{year}{2019}), \bibinfo{pages}{230--239}.
\newblock
\showISBNx{9781450361255}
\urldef\tempurl%
\url{https://doi.org/10.1145/3287560.3287576}
\showDOI{\tempurl}


\bibitem[Mitchell et~al\mbox{.}(2021)]%
        {mitchell2021algorithmic}
\bibfield{author}{\bibinfo{person}{Shira Mitchell}, \bibinfo{person}{Eric
  Potash}, \bibinfo{person}{Solon Barocas}, \bibinfo{person}{Alexander
  D'Amour}, {and} \bibinfo{person}{Kristian Lum}.}
  \bibinfo{year}{2021}\natexlab{}.
\newblock \showarticletitle{{Algorithmic Fairness: Choices, Assumptions, and
  Definitions}}.
\newblock \bibinfo{journal}{\emph{Annual Review of Statistics and Its
  Application}} \bibinfo{volume}{8}, \bibinfo{number}{1} (\bibinfo{date}{3}
  \bibinfo{year}{2021}), \bibinfo{pages}{141--163}.
\newblock
\showISSN{2326-8298}
\urldef\tempurl%
\url{https://doi.org/10.1146/annurev-statistics-042720-125902}
\showDOI{\tempurl}


\bibitem[Mouzannar et~al\mbox{.}(2019)]%
        {Mouzannar2019FromEquality}
\bibfield{author}{\bibinfo{person}{Hussein Mouzannar},
  \bibinfo{person}{Mesrob~I. Ohannessian}, {and} \bibinfo{person}{Nathan
  Srebro}.} \bibinfo{year}{2019}\natexlab{}.
\newblock \showarticletitle{{From fair decision making to social equality}}.
\newblock \bibinfo{journal}{\emph{FAT* 2019 - Proceedings of the 2019
  Conference on Fairness, Accountability, and Transparency}}
  (\bibinfo{year}{2019}), \bibinfo{pages}{359--368}.
\newblock
\showISBNx{9781450361255}
\urldef\tempurl%
\url{https://doi.org/10.1145/3287560.3287599}
\showDOI{\tempurl}


\bibitem[Obermajer et~al\mbox{.}(1983)]%
        {NO-RM-JL-RPE-PK:11}
\bibfield{author}{\bibinfo{person}{Nataša Obermajer},
  \bibinfo{person}{Ravikumar Muthuswamy}, \bibinfo{person}{Jamie Lesnock},
  \bibinfo{person}{Robert~P Edwards}, {and} \bibinfo{person}{Pawel Kalinski}.}
  \bibinfo{year}{1983}\natexlab{}.
\newblock \showarticletitle{{Positive feedback between PGE{\$}{\_}2{\$} and
  COX2 redirects the differentiation of human dendritic cells toward stable
  myeloid-derived suppressor cells}}.
\newblock \bibinfo{journal}{\emph{Immunobiology}} \bibinfo{volume}{119},
  \bibinfo{number}{20} (\bibinfo{year}{1983}).
\newblock
\urldef\tempurl%
\url{https://doi.org/10.1182/blood-2011-07-365825}
\showDOI{\tempurl}


\bibitem[Olteanu et~al\mbox{.}(2019)]%
        {olteanu2019socialdata}
\bibfield{author}{\bibinfo{person}{Alexandra Olteanu}, \bibinfo{person}{Carlos
  Castillo}, \bibinfo{person}{Fernando Diaz}, {and} \bibinfo{person}{Emre
  Kıcıman}.} \bibinfo{year}{2019}\natexlab{}.
\newblock \showarticletitle{{Social Data: Biases, Methodological Pitfalls, and
  Ethical Boundaries}}.
\newblock \bibinfo{journal}{\emph{Frontiers in Big Data}}  \bibinfo{volume}{2}
  (\bibinfo{date}{7} \bibinfo{year}{2019}).
\newblock
\showISSN{2624-909X}
\urldef\tempurl%
\url{https://doi.org/10.3389/fdata.2019.00013}
\showDOI{\tempurl}


\bibitem[O'neil(2017)]%
        {oneil2017weapons}
\bibfield{author}{\bibinfo{person}{Cathy O'neil}.}
  \bibinfo{year}{2017}\natexlab{}.
\newblock \bibinfo{booktitle}{\emph{{Weapons of math destruction: How big data
  increases inequality and threatens democracy}}}.
\newblock \bibinfo{publisher}{Crown}.
\newblock


\bibitem[Perdomo et~al\mbox{.}(2020)]%
        {Perdomo2020PerformativePrediction}
\bibfield{author}{\bibinfo{person}{Juan~C. Perdomo}, \bibinfo{person}{Tijana
  Zrnic}, \bibinfo{person}{Celestine Mendler-Dunner}, {and}
  \bibinfo{person}{Moritz Hardt}.} \bibinfo{year}{2020}\natexlab{}.
\newblock \showarticletitle{{Performative prediction}}.
\newblock \bibinfo{journal}{\emph{37th International Conference on Machine
  Learning, ICML 2020}}  \bibinfo{volume}{PartF16814} (\bibinfo{year}{2020}),
  \bibinfo{pages}{7555--7565}.
\newblock
\showISBNx{9781713821120}


\bibitem[Perra and Rocha(2019)]%
        {perra2019modelling}
\bibfield{author}{\bibinfo{person}{Nicola Perra} {and} \bibinfo{person}{Luis
  E~C Rocha}.} \bibinfo{year}{2019}\natexlab{}.
\newblock \showarticletitle{{Modelling opinion dynamics in the age of
  algorithmic personalisation}}.
\newblock \bibinfo{journal}{\emph{Scientific reports}} \bibinfo{volume}{9},
  \bibinfo{number}{1} (\bibinfo{year}{2019}), \bibinfo{pages}{1--11}.
\newblock


\bibitem[Pessach and Shmueli(2022)]%
        {pessach2022survey}
\bibfield{author}{\bibinfo{person}{Dana Pessach} {and} \bibinfo{person}{Erez
  Shmueli}.} \bibinfo{year}{2022}\natexlab{}.
\newblock \showarticletitle{{A Review on Fairness in Machine Learning}}.
\newblock \bibinfo{journal}{\emph{ACM Comput. Surv.}} \bibinfo{volume}{55},
  \bibinfo{number}{3} (\bibinfo{date}{2} \bibinfo{year}{2022}).
\newblock
\showISSN{0360-0300}
\urldef\tempurl%
\url{https://doi.org/10.1145/3494672}
\showDOI{\tempurl}


\bibitem[Ramaprasad(1983)]%
        {AR:83}
\bibfield{author}{\bibinfo{person}{Arkalgud Ramaprasad}.}
  \bibinfo{year}{1983}\natexlab{}.
\newblock \showarticletitle{{On the definition of feedback}}.
\newblock \bibinfo{journal}{\emph{Journal of the Society for General Systems
  Research}} \bibinfo{volume}{28}, \bibinfo{number}{1} (\bibinfo{year}{1983}).
\newblock
\urldef\tempurl%
\url{https://doi.org/10.1002/bs.3830280103}
\showDOI{\tempurl}


\bibitem[Rossi et~al\mbox{.}(2021)]%
        {9516926}
\bibfield{author}{\bibinfo{person}{Wilbert~Samuel Rossi},
  \bibinfo{person}{Jan~Willem Polderman}, {and} \bibinfo{person}{Paolo
  Frasca}.} \bibinfo{year}{2021}\natexlab{}.
\newblock \showarticletitle{{The closed loop between opinion formation and
  personalised recommendations}}.
\newblock \bibinfo{journal}{\emph{IEEE Transactions on Control of Network
  Systems}} (\bibinfo{year}{2021}), \bibinfo{pages}{1}.
\newblock
\urldef\tempurl%
\url{https://doi.org/10.1109/TCNS.2021.3105616}
\showDOI{\tempurl}


\bibitem[Simonite(2015)]%
        {simonite2015probing}
\bibfield{author}{\bibinfo{person}{Tom Simonite}.}
  \bibinfo{year}{2015}\natexlab{}.
\newblock \showarticletitle{{Probing the dark side of google’s ad-targeting
  system}}.
\newblock \bibinfo{journal}{\emph{MIT Technology Review}}
  (\bibinfo{year}{2015}).
\newblock


\bibitem[Sinha et~al\mbox{.}(2016)]%
        {NIPS2016_962e56a8}
\bibfield{author}{\bibinfo{person}{Ayan Sinha}, \bibinfo{person}{David~F
  Gleich}, {and} \bibinfo{person}{Karthik Ramani}.}
  \bibinfo{year}{2016}\natexlab{}.
\newblock \showarticletitle{{Deconvolving Feedback Loops in Recommender
  Systems}}. In \bibinfo{booktitle}{\emph{Advances in Neural Information
  Processing Systems}}, \bibfield{editor}{\bibinfo{person}{D~Lee},
  \bibinfo{person}{M~Sugiyama}, \bibinfo{person}{U~Luxburg},
  \bibinfo{person}{I~Guyon}, {and} \bibinfo{person}{R~Garnett}} (Eds.),
  Vol.~\bibinfo{volume}{29}. \bibinfo{publisher}{Curran Associates, Inc.}
\newblock
\urldef\tempurl%
\url{https://proceedings.neurips.cc/paper/2016/file/962e56a8a0b0420d87272a682bfd1e53-Paper.pdf}
\showURL{%
\tempurl}


\bibitem[Sun(2022)]%
        {AliciaYiSun2022PhDThesis}
\bibfield{author}{\bibinfo{person}{Yi Sun}.} \bibinfo{year}{2022}\natexlab{}.
\newblock \emph{\bibinfo{title}{{Algorithmic Fairness in Sequential Decision
  Making}}}.
\newblock \bibinfo{thesistype}{Ph.\,D. Dissertation}.
\newblock


\bibitem[Sun et~al\mbox{.}(2022)]%
        {sunbackfire}
\bibfield{author}{\bibinfo{person}{Yi Sun}, \bibinfo{person}{Alfredo
  Cuesta-Infante}, {and} \bibinfo{person}{Kalyan Veeramachaneni}.}
  \bibinfo{year}{2022}\natexlab{}.
\newblock \showarticletitle{{The Backfire Effects of Fairness Constraints}}.
\newblock \bibinfo{journal}{\emph{ICML 2022 Workshop on Responsible Decision
  Making in Dynamic Environments}} (\bibinfo{year}{2022}).
\newblock
\urldef\tempurl%
\url{https://responsibledecisionmaking.github.io/assets/pdf/papers/44.pdf}
\showURL{%
\tempurl}


\bibitem[Suresh and Guttag(2021)]%
        {10.1145/3465416.3483305}
\bibfield{author}{\bibinfo{person}{Harini Suresh} {and} \bibinfo{person}{John
  Guttag}.} \bibinfo{year}{2021}\natexlab{}.
\newblock \showarticletitle{{A Framework for Understanding Sources of Harm
  throughout the Machine Learning Life Cycle}}. In
  \bibinfo{booktitle}{\emph{Equity and Access in Algorithms, Mechanisms, and
  Optimization}} \emph{(\bibinfo{series}{EAAMO '21})}.
  \bibinfo{publisher}{Association for Computing Machinery},
  \bibinfo{address}{New York, NY, USA}.
\newblock
\showISBNx{9781450385534}
\urldef\tempurl%
\url{https://doi.org/10.1145/3465416.3483305}
\showDOI{\tempurl}


\bibitem[Tsirtsis et~al\mbox{.}(2019)]%
        {Tsirtsis2019OptimalBehavior}
\bibfield{author}{\bibinfo{person}{Stratis Tsirtsis}, \bibinfo{person}{Behzad
  Tabibian}, \bibinfo{person}{Moein Khajehnejad}, \bibinfo{person}{Adish
  Singla}, \bibinfo{person}{Bernhard Sch{\"{o}}lkopf}, {and}
  \bibinfo{person}{Manuel Gomez-Rodriguez}.} \bibinfo{year}{2019}\natexlab{}.
\newblock \showarticletitle{{Optimal Decision Making Under Strategic
  Behavior}}.
\newblock  (\bibinfo{year}{2019}).
\newblock
\urldef\tempurl%
\url{http://arxiv.org/abs/1905.09239}
\showURL{%
\tempurl}


\bibitem[van Giffen et~al\mbox{.}(2022)]%
        {VanGiffen2022biases}
\bibfield{author}{\bibinfo{person}{Benjamin van Giffen},
  \bibinfo{person}{Dennis Herhausen}, {and} \bibinfo{person}{Tobias Fahse}.}
  \bibinfo{year}{2022}\natexlab{}.
\newblock \showarticletitle{{Overcoming the pitfalls and perils of algorithms:
  A classification of machine learning biases and mitigation methods}}.
\newblock \bibinfo{journal}{\emph{Journal of Business Research}}
  \bibinfo{volume}{144} (\bibinfo{year}{2022}), \bibinfo{pages}{93--106}.
\newblock
\showISSN{0148-2963}
\urldef\tempurl%
\url{https://doi.org/10.1016/j.jbusres.2022.01.076}
\showDOI{\tempurl}


\bibitem[Vorobeychik and Kantarcioglu(2018)]%
        {Vorobeychik2018AdversarialMachineLearning}
\bibfield{author}{\bibinfo{person}{Yevgeniy Vorobeychik} {and}
  \bibinfo{person}{Murat Kantarcioglu}.} \bibinfo{year}{2018}\natexlab{}.
\newblock \bibinfo{booktitle}{\emph{{Adversarial Machine Learning}}}.
\newblock \bibinfo{publisher}{Springer International Publishing},
  \bibinfo{address}{Cham}.
\newblock
\showISBNx{978-3-031-00452-0}
\urldef\tempurl%
\url{https://doi.org/10.1007/978-3-031-01580-9}
\showDOI{\tempurl}


\bibitem[Webb et~al\mbox{.}(2016)]%
        {Webb2016Characterizing}
\bibfield{author}{\bibinfo{person}{Geoffrey~I Webb}, \bibinfo{person}{Roy
  Hyde}, \bibinfo{person}{Hong Cao}, \bibinfo{person}{Hai~Long Nguyen}, {and}
  \bibinfo{person}{Francois Petitjean}.} \bibinfo{year}{2016}\natexlab{}.
\newblock \showarticletitle{{Characterizing concept drift}}.
\newblock \bibinfo{journal}{\emph{Data Mining and Knowledge Discovery}}
  \bibinfo{volume}{30}, \bibinfo{number}{4} (\bibinfo{year}{2016}),
  \bibinfo{pages}{964--994}.
\newblock
\showISSN{1573-756X}
\urldef\tempurl%
\url{https://doi.org/10.1007/s10618-015-0448-4}
\showDOI{\tempurl}


\bibitem[Xu et~al\mbox{.}(2021)]%
        {Xu2021RobustOrFair}
\bibfield{author}{\bibinfo{person}{Han Xu}, \bibinfo{person}{Xiaorui Liu},
  \bibinfo{person}{Yaxin Li}, \bibinfo{person}{Anil Jain}, {and}
  \bibinfo{person}{Jiliang Tang}.} \bibinfo{year}{2021}\natexlab{}.
\newblock \showarticletitle{{To be Robust or to be Fair: Towards Fairness in
  Adversarial Training}}. In \bibinfo{booktitle}{\emph{Proceedings of the 38th
  International Conference on Machine Learning}}
  \emph{(\bibinfo{series}{Proceedings of Machine Learning Research},
  Vol.~\bibinfo{volume}{139})}, \bibfield{editor}{\bibinfo{person}{Marina
  Meila} {and} \bibinfo{person}{Tong Zhang}} (Eds.). \bibinfo{publisher}{PMLR},
  \bibinfo{pages}{11492--11501}.
\newblock
\urldef\tempurl%
\url{https://proceedings.mlr.press/v139/xu21b.html}
\showURL{%
\tempurl}


\bibitem[Zeigler et~al\mbox{.}(2000)]%
        {zeigler2000theory}
\bibfield{author}{\bibinfo{person}{Bernard~P Zeigler}, \bibinfo{person}{Tag~Gon
  Kim}, {and} \bibinfo{person}{Herbert Praehofer}.}
  \bibinfo{year}{2000}\natexlab{}.
\newblock \bibinfo{booktitle}{\emph{{Theory of modeling and simulation}}}.
\newblock \bibinfo{publisher}{Academic press}.
\newblock


\bibitem[Zhang et~al\mbox{.}(2020a)]%
        {Zhang2020Long-TermLearning}
\bibfield{author}{\bibinfo{person}{Xueru Zhang},
  \bibinfo{person}{Mohammad~Mahdi Khalili}, {and} \bibinfo{person}{Mingyan
  Liu}.} \bibinfo{year}{2020}\natexlab{a}.
\newblock \showarticletitle{{Long-Term Impacts of Fair Machine Learning}}.
\newblock \bibinfo{journal}{\emph{Ergonomics in Design}} \bibinfo{volume}{28},
  \bibinfo{number}{3} (\bibinfo{year}{2020}), \bibinfo{pages}{7--11}.
\newblock
\showISSN{10648046}
\urldef\tempurl%
\url{https://doi.org/10.1177/1064804619884160}
\showDOI{\tempurl}


\bibitem[Zhang et~al\mbox{.}(2019)]%
        {Zhang2019GroupFairness}
\bibfield{author}{\bibinfo{person}{Xueru Zhang},
  \bibinfo{person}{Mohammad~Mahdi Khalili}, \bibinfo{person}{Cem Tekin}, {and}
  \bibinfo{person}{Mingyan Liu}.} \bibinfo{year}{2019}\natexlab{}.
\newblock \showarticletitle{{Group retention when using machine learning in
  sequential decision making: The interplay between user dynamics and
  fairness}}.
\newblock \bibinfo{journal}{\emph{Advances in Neural Information Processing
  Systems}} \bibinfo{volume}{32}, \bibinfo{number}{NeurIPS}
  (\bibinfo{year}{2019}).
\newblock
\showISSN{10495258}


\bibitem[Zhang and Liu(2021)]%
        {Zhang2021FairnessSurvey}
\bibfield{author}{\bibinfo{person}{Xueru Zhang} {and} \bibinfo{person}{Mingyan
  Liu}.} \bibinfo{year}{2021}\natexlab{}.
\newblock \showarticletitle{{Fairness in Learning-Based Sequential Decision
  Algorithms: A Survey}}.
\newblock \bibinfo{journal}{\emph{Studies in Systems, Decision and Control}}
  \bibinfo{volume}{325} (\bibinfo{year}{2021}), \bibinfo{pages}{525--555}.
\newblock
\showISSN{21984190}
\urldef\tempurl%
\url{https://doi.org/10.1007/978-3-030-60990-0{\_}18}
\showDOI{\tempurl}


\bibitem[Zhang et~al\mbox{.}(2020b)]%
        {Zhang2020HowQualification}
\bibfield{author}{\bibinfo{person}{Xueru Zhang}, \bibinfo{person}{Ruibo Tu},
  \bibinfo{person}{Yang Liu}, \bibinfo{person}{Mingyan Liu},
  \bibinfo{person}{Hedvig Kjellstr{\"{o}}m}, \bibinfo{person}{Kun Zhang}, {and}
  \bibinfo{person}{Cheng Zhang}.} \bibinfo{year}{2020}\natexlab{b}.
\newblock \showarticletitle{{How do fair decisions fare in long-term
  qualification?}}
\newblock \bibinfo{journal}{\emph{Advances in Neural Information Processing
  Systems}} \bibinfo{volume}{2020-Decem}, \bibinfo{number}{NeurIPS}
  (\bibinfo{year}{2020}), \bibinfo{pages}{1--13}.
\newblock
\showISSN{10495258}


\bibitem[Zhou and Doyle(1998)]%
        {zhou1998essentials}
\bibfield{author}{\bibinfo{person}{Kemin Zhou} {and}
  \bibinfo{person}{John~Comstock Doyle}.} \bibinfo{year}{1998}\natexlab{}.
\newblock \bibinfo{booktitle}{\emph{Essentials of robust control}}.
  Vol.~\bibinfo{volume}{104}.
\newblock \bibinfo{publisher}{Prentice hall Upper Saddle River, NJ}.
\newblock


\end{thebibliography}

\appendix

\section{Literature search process}
\label{app:literature_search}

We conducted a literature search to identify relevant articles for our analysis.
We started by performing a forward and backward search using an initially small set of important papers~\cite{Ensign2018RunawayPolicing,DAmour2020FairnessStudies,Liu2018Delayed_Long_version,Zhang2021FairnessSurvey}.
Then, we performed a keyword search on the ACM Digital Library and google scholar using various combinations of the following keyword: ``feedback loop,'' ``algorithmic bias,'' ``algorithmic fairness,'' ``machine learning,'' and ``feedback.''
We performed the classification of existing works in three stages:
First, we skimmed each paper and added it to a list of potentially relevant papers if it contained some mention of feedback effects and ML -- this list consisted of 75 papers.
Next, we looked at each paper in more detail and only included it in the final base of literature to be considered if it investigates feedback loops as part of ML-based decision-making systems.
Finally, looked at each description of feedback loops in those papers and mapped it to our conceptualization of feedback loops.
This resulted in the 24 papers listed in Table~\ref{table:literature} and, through an iterative process, ultimately also in Figure~\ref{fig:feedback-loops} presented in Section~\ref{ssec:Feedback-loops}.

\section{Open- and closed-loop dynamical systems}
\label{app:Open_and_closed_loop_dynamical_systems}

\begin{figure}[!htb]
\centering
\begin{subfigure}[b]{0.45\textwidth}
\centering
\includegraphics[width=\textwidth]{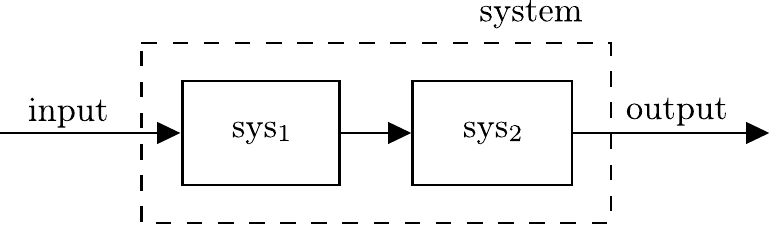}
\vspace*{5mm}
\caption{Open-loop}
\label{fig:systems-interconnection_open}
\end{subfigure}
\hfil
\begin{subfigure}[b]{0.5\textwidth}
\centering
\includegraphics[width=\textwidth]{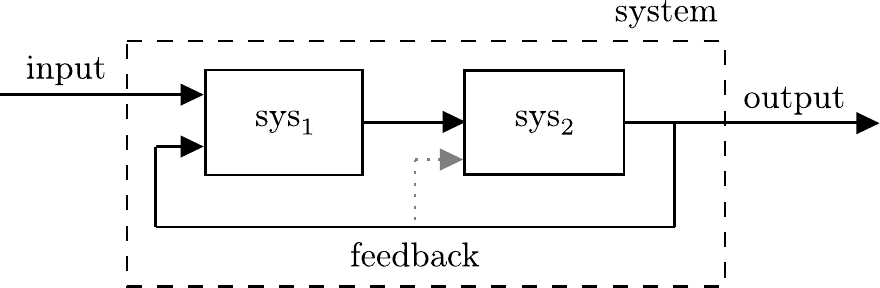}
\vspace*{5mm}
\caption{Closed-loop}
\label{fig:systems-interconnection_closed}
\end{subfigure}
\caption{Open- and closed-loop dynamical systems
}
\label{fig:systems-interconnection}
\end{figure}

In Fig.~\ref{fig:systems-interconnection}, we provide a simple visualization of the two types of \emph{series} and \emph{feedback interconnections}.
A \emph{series interconnection} (Fig.~\ref{fig:systems-interconnection_open}) composes two systems into an \emph{open-loop system}.
A \emph{feedback interconnection} (Fig.~\ref{fig:systems-interconnection_closed}) composes them into a \emph{closed-loop system}: the output is injected back as an input to one (or more) of the components, creating a \emph{feedback loop}.
Unlike their open-loop counterpart, closed-loop systems are not straightforward to predict from their components and require specialized techniques to analyze.

The interconnections of systems are systems themselves, and a model for the interconnected system can be derived from the models of the individual subsystems that compose the interconnection. 
This derivation is straightforward in the case of series interconnection, but it becomes significantly more involved in the case of feedback interconnection.
\saverio{Only for special classes of systems, for example, systems where the input-output relation of each subsystem is linear, a model for the resulting interconnected system can be derived analytically.
For general dynamical systems, a tractable direct derivation is typically very difficult, and numerical methods (including simulations) come to help.
}

\section{ML-Based Decision-Making Pipeline}
\label{app:detailed_ML_pipeline}

\begin{figure}[!ht]
\includegraphics[width=1\textwidth]{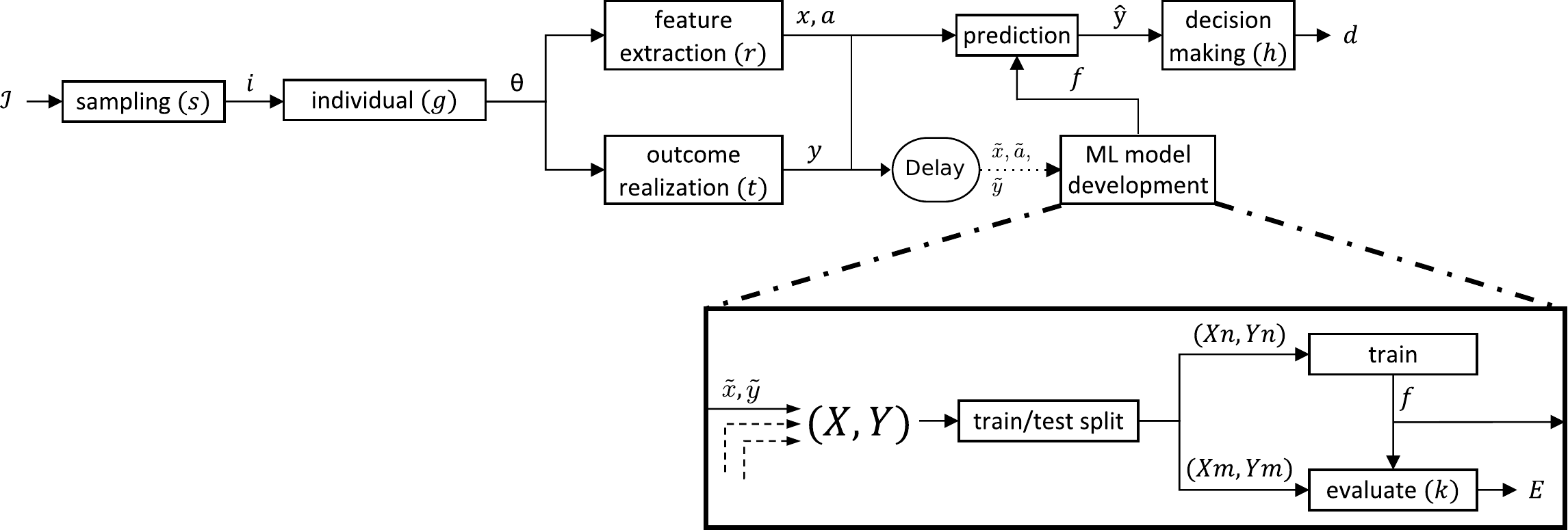}
\centering
\caption{The detailed ML-based decision-making pipeline}
\label{fig:pipeline_MLmodel}
\end{figure}

Fig.~\ref{fig:pipeline_MLmodel} visualizes a more detailed ML pipeline, also zooming into the ML model development process.
It shows that once observed, an individual's feature label pair $x,y$ can end up in a sample $(X,Y)$ that is used to (re)train and evaluate a predictor.
This sample is split into training data $(X_n,Y_n)$ and testing data $(X_m,Y_m)$.
The training data is used to learn a function $f : x\rightarrow \hat{y}$.
This learned function is evaluated using the test data, which outputs some evaluation metrics $E$ that are computed using a function $k : f,X_m,Y_m \rightarrow E$.
Finally, $f$ is used to predict the outcome of previously unseen feature values in the next iteration.

Notice that this is an example of a potential extension of the feedback loop classification presented in Section~\ref{ssec:Feedback-loops}: the ML model feedback loop could also be split up into an ML model training feedback and an ML model evaluation feedback loop.
However, once a feature label pair is added to $(X,Y)$, it is usually assigned randomly to either the training or the evaluation data (or even to both in the process of cross-validation), which is why we chose to use the umbrella term ML model feedback loop.

\section{Addendum to case study on recommender systems}
\label{app:case_study_addendum}

\begin{figure}[!htb]
\centering
\begin{tikzpicture}

\definecolor{color0}{rgb}{0.843137254901961,0.0980392156862745,0.109803921568627}
\definecolor{color1}{rgb}{0.172549019607843,0.482352941176471,0.713725490196078}

\begin{axis}[
height=6cm,
width=8cm,
legend cell align={left},
legend style={fill opacity=0.8, draw opacity=1, text opacity=1, draw=white!80!black},
tick align=outside,
tick pos=left,
x grid style={white!69.0196078431373!black},
xmin=0, xmax=1,
xtick style={color=black},
y grid style={white!69.0196078431373!black},
ymin=0, ymax=68.25,
ylabel={Number of Users},
xlabel={$\theta$},
ytick style={color=black},
]
\draw[draw=none,fill=color0,fill opacity=0.8] (axis cs:0.192781920195074,0) rectangle (axis cs:0.233046067001822,2);
\addlegendimage{ybar,ybar legend,draw=none,fill=color0,fill opacity=0.8};
\addlegendentry{Group 1}

\draw[draw=none,fill=color0,fill opacity=0.8] (axis cs:0.233046067001822,0) rectangle (axis cs:0.273310213808569,1);
\draw[draw=none,fill=color0,fill opacity=0.8] (axis cs:0.273310213808569,0) rectangle (axis cs:0.313574360615316,0);
\draw[draw=none,fill=color0,fill opacity=0.8] (axis cs:0.313574360615316,0) rectangle (axis cs:0.353838507422063,3);
\draw[draw=none,fill=color0,fill opacity=0.8] (axis cs:0.353838507422063,0) rectangle (axis cs:0.39410265422881,7);
\draw[draw=none,fill=color0,fill opacity=0.8] (axis cs:0.39410265422881,0) rectangle (axis cs:0.434366801035558,3);
\draw[draw=none,fill=color0,fill opacity=0.8] (axis cs:0.434366801035558,0) rectangle (axis cs:0.474630947842305,10);
\draw[draw=none,fill=color0,fill opacity=0.8] (axis cs:0.474630947842305,0) rectangle (axis cs:0.514895094649052,22);
\draw[draw=none,fill=color0,fill opacity=0.8] (axis cs:0.514895094649052,0) rectangle (axis cs:0.555159241455799,37);
\draw[draw=none,fill=color0,fill opacity=0.8] (axis cs:0.555159241455799,0) rectangle (axis cs:0.595423388262546,37);
\draw[draw=none,fill=color0,fill opacity=0.8] (axis cs:0.595423388262546,0) rectangle (axis cs:0.635687535069294,50);
\draw[draw=none,fill=color0,fill opacity=0.8] (axis cs:0.635687535069294,0) rectangle (axis cs:0.675951681876041,65);
\draw[draw=none,fill=color0,fill opacity=0.8] (axis cs:0.675951681876041,0) rectangle (axis cs:0.716215828682788,43);
\draw[draw=none,fill=color0,fill opacity=0.8] (axis cs:0.716215828682788,0) rectangle (axis cs:0.756479975489535,49);
\draw[draw=none,fill=color0,fill opacity=0.8] (axis cs:0.756479975489535,0) rectangle (axis cs:0.796744122296283,42);
\draw[draw=none,fill=color0,fill opacity=0.8] (axis cs:0.796744122296283,0) rectangle (axis cs:0.83700826910303,37);
\draw[draw=none,fill=color0,fill opacity=0.8] (axis cs:0.83700826910303,0) rectangle (axis cs:0.877272415909777,39);
\draw[draw=none,fill=color0,fill opacity=0.8] (axis cs:0.877272415909777,0) rectangle (axis cs:0.917536562716524,22);
\draw[draw=none,fill=color0,fill opacity=0.8] (axis cs:0.917536562716524,0) rectangle (axis cs:0.957800709523271,20);
\draw[draw=none,fill=color0,fill opacity=0.8] (axis cs:0.957800709523271,0) rectangle (axis cs:0.998064856330018,7);
\draw[draw=none,fill=color1,fill opacity=0.8] (axis cs:0.0117855441296537,0) rectangle (axis cs:0.0491027995967672,6);
\addlegendimage{ybar,ybar legend,draw=none,fill=color1,fill opacity=0.8};
\addlegendentry{Group 2}

\draw[draw=none,fill=color1,fill opacity=0.8] (axis cs:0.0491027995967672,0) rectangle (axis cs:0.0864200550638806,26);
\draw[draw=none,fill=color1,fill opacity=0.8] (axis cs:0.0864200550638806,0) rectangle (axis cs:0.123737310530994,31);
\draw[draw=none,fill=color1,fill opacity=0.8] (axis cs:0.123737310530994,0) rectangle (axis cs:0.161054565998108,33);
\draw[draw=none,fill=color1,fill opacity=0.8] (axis cs:0.161054565998108,0) rectangle (axis cs:0.198371821465221,23);
\draw[draw=none,fill=color1,fill opacity=0.8] (axis cs:0.198371821465221,0) rectangle (axis cs:0.235689076932335,42);
\draw[draw=none,fill=color1,fill opacity=0.8] (axis cs:0.235689076932335,0) rectangle (axis cs:0.273006332399448,45);
\draw[draw=none,fill=color1,fill opacity=0.8] (axis cs:0.273006332399448,0) rectangle (axis cs:0.310323587866561,45);
\draw[draw=none,fill=color1,fill opacity=0.8] (axis cs:0.310323587866562,0) rectangle (axis cs:0.347640843333675,47);
\draw[draw=none,fill=color1,fill opacity=0.8] (axis cs:0.347640843333675,0) rectangle (axis cs:0.384958098800788,41);
\draw[draw=none,fill=color1,fill opacity=0.8] (axis cs:0.384958098800788,0) rectangle (axis cs:0.422275354267902,42);
\draw[draw=none,fill=color1,fill opacity=0.8] (axis cs:0.422275354267902,0) rectangle (axis cs:0.459592609735015,31);
\draw[draw=none,fill=color1,fill opacity=0.8] (axis cs:0.459592609735015,0) rectangle (axis cs:0.496909865202129,28);
\draw[draw=none,fill=color1,fill opacity=0.8] (axis cs:0.496909865202129,0) rectangle (axis cs:0.534227120669242,24);
\draw[draw=none,fill=color1,fill opacity=0.8] (axis cs:0.534227120669242,0) rectangle (axis cs:0.571544376136356,19);
\draw[draw=none,fill=color1,fill opacity=0.8] (axis cs:0.571544376136356,0) rectangle (axis cs:0.608861631603469,9);
\draw[draw=none,fill=color1,fill opacity=0.8] (axis cs:0.608861631603469,0) rectangle (axis cs:0.646178887070583,5);
\draw[draw=none,fill=color1,fill opacity=0.8] (axis cs:0.646178887070583,0) rectangle (axis cs:0.683496142537696,2);
\draw[draw=none,fill=color1,fill opacity=0.8] (axis cs:0.683496142537696,0) rectangle (axis cs:0.72081339800481,0);
\draw[draw=none,fill=color1,fill opacity=0.8] (axis cs:0.72081339800481,0) rectangle (axis cs:0.758130653471923,5);
\end{axis}

\end{tikzpicture}
\caption{Initial empirical $\theta$ distribution used in all the examples except the one on feature feedback loop.}
\label{fig:initial_theta_distribution}
\end{figure}
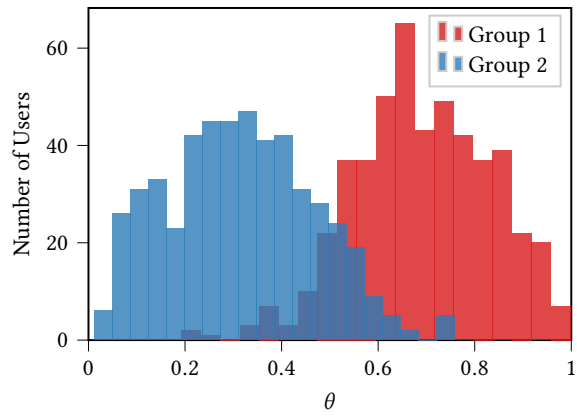

Fig.~\ref{fig:initial_theta_distribution} shows the initial empirical $\theta$ distribution used in all the examples in Section~\ref{sec:example} except the one on feature feedback loop.
For the feature feedback loop, the two initial $\theta$ distributions for groups G1 and G2 have the same mean value of $\left(\mu_{\theta,G1}=\mu_{\theta,G2}=0.5 \right)$, i.e., in this case, there is no group-level difference between the platform users' interests.

\end{document}